\documentclass[twocolumn]{aastex631}

\usepackage{multirow}

\makeatletter
\newcommand*{\linktocite}[2]{%
  \hyper@natlinkstart{#1}#2\hyper@natlinkend}
\makeatother

\shortauthors{Lin et al.}
\graphicspath{{./}{figures/}}
\begin{document}

\title{H$_2$-dominated Atmosphere as an Indicator of Second-generation Rocky White Dwarf Exoplanets}

\author[0000-0003-0525-9647]{Zifan Lin}
\affiliation{Department of Earth, Atmospheric, and Planetary Sciences, Massachusetts Institute of Technology, 77 Massachusetts Avenue, Cambridge, MA 02139, USA}

\author{Sara Seager}
\affiliation{Department of Earth, Atmospheric, and Planetary Sciences, Massachusetts Institute of Technology, 77 Massachusetts Avenue, Cambridge, MA 02139, USA}
\affiliation{Department of Physics, Massachusetts Institute of Technology, 77 Massachusetts Avenue, Cambridge, MA 02139, USA}
\affiliation{Department of Aeronautics and Astronautics, Massachusetts Institute of Technology, 77 Massachusetts Avenue, Cambridge, MA 02139, USA}

\author[0000-0002-5147-9053]{Sukrit Ranjan}
\affiliation{Northwestern University, Center for Interdisciplinary Exploration and Research in Astrophysics, Evanston, 60201, USA}
\affiliation{Northwestern University, Department of Astronomy \& Astrophysics, Evanston, 60201, USA}
\affiliation{Blue Marble Space Institute of Science, Seattle, 98154, USA}

\author[0000-0002-3868-2129]{Thea Kozakis}
\affiliation{DTU Space, National Space Institute, Technical University of Denmark, Elektrovej 328, DK-2800 Kgs. Lyngby, Denmark}

\author[0000-0002-0436-1802]{Lisa Kaltenegger}
\affiliation{Carl Sagan Institute and Department of Astronomy, Cornell University, Ithaca, NY 14853, USA}



%% 158 / 250 word limit
\begin{abstract}

Following the discovery of the first exoplanet candidate transiting a white dwarf (WD), a ``white dwarf opportunity" for characterizing the atmospheres of terrestrial exoplanets around WDs is emerging. Large planet-to-star size ratios and hence large transit depths make transiting WD exoplanets favorable targets for transmission spectroscopy -- conclusive detection of spectral features on an Earth-like planet transiting a close-by WD can be achieved within a medium James Webb Space Telescope (JWST) program. Despite the apparently promising opportunity, however, the post-main sequence (MS) evolutionary history of a first-generation WD exoplanet has never been incorporated in atmospheric modeling. Furthermore, second-generation planets formed in WD debris disks have never been studied from a photochemical perspective. We demonstrate that transmission spectroscopy can identify a second-generation rocky WD exoplanet with a thick ($\sim1$ bar) H$_2$-dominated atmosphere. In addition, we can infer outgassing activities of a WD exoplanet based on its transmission spectra and test photochemical runaway by studying CH$_4$ buildup.

\end{abstract}



%%======================================================================================
%%  INTRODUCTION
%%======================================================================================

\section{Introduction} \label{sec:intro}
An exciting opportunity for characterizing the atmospheres of terrestrial exoplanets transiting WDs is emerging. The first planet candidate transiting a WD was discovered in the WD 1856+534 system \citep{vanderburg_giant_2020}, followed by a recent microlensing detection of a gas giant in a Jupiter-like orbit around a WD \citep{Blackman_2021Natur.598..272B}. \linktocite{kaltenegger_white_2020}{Kaltenegger, MacDonald et al.\ (2020)} explored the possibility of observing transiting Earth-like WD planets with JWST and described a ``white dwarf opportunity'' of detecting biosignature gases on such planets. Due to large planet-to-star radius ratio, WD exoplanets have much larger transit depths compared to planets around MS hosts, making them favorable targets for transmission spectroscopy \citep[e.g.,][]{agol_transit_2011, loeb_detecting_2013}. For a hypothetical Earth-sized planet with Earth-like atmosphere transiting WD 1856+534, JWST can detect H$_2$O and CO$_2$ with just a few transits and detect biosignature gases such as the O$_3$ + CH$_4$ pair in 25 transits (\linktocite{kaltenegger_white_2020}{Kaltenegger, MacDonald et al.\ 2020}). For comparison, JWST would struggle to detect the O$_3$ + CH$_4$ biosignature pair on a terrestrial planet orbiting a M dwarf such as TRAPPIST-1e even with 100 transits \citep[e.g.,][]{lin_differentiating_2021}.

There have been some pioneering works on photochemical modeling of WD exoplanets, but those works are naturally limited in the parameter space explored. \cite{kozakis_uv_2018, kozakis_high-resolution_2020} modeled Earth-mass planets orbiting WDs with N$_2$-dominated atmospheres, modern Earth-like O$_2$ and CO$_2$ concentrations, and modern-Earth like outgassing rates for key spectral species including CH$_4$ and N$_2$O. \linktocite{kaltenegger_white_2020}{Kaltenegger, MacDonald et al.\ (2020)} analyzed the detectability of spectral signatures by JWST for Earth-like rocky WD exoplanets, based on the models developed by \cite{kozakis_uv_2018, kozakis_high-resolution_2020}.

%%% STELLAR SPECTRA ================
\begin{figure*}[ht!]
\plotone{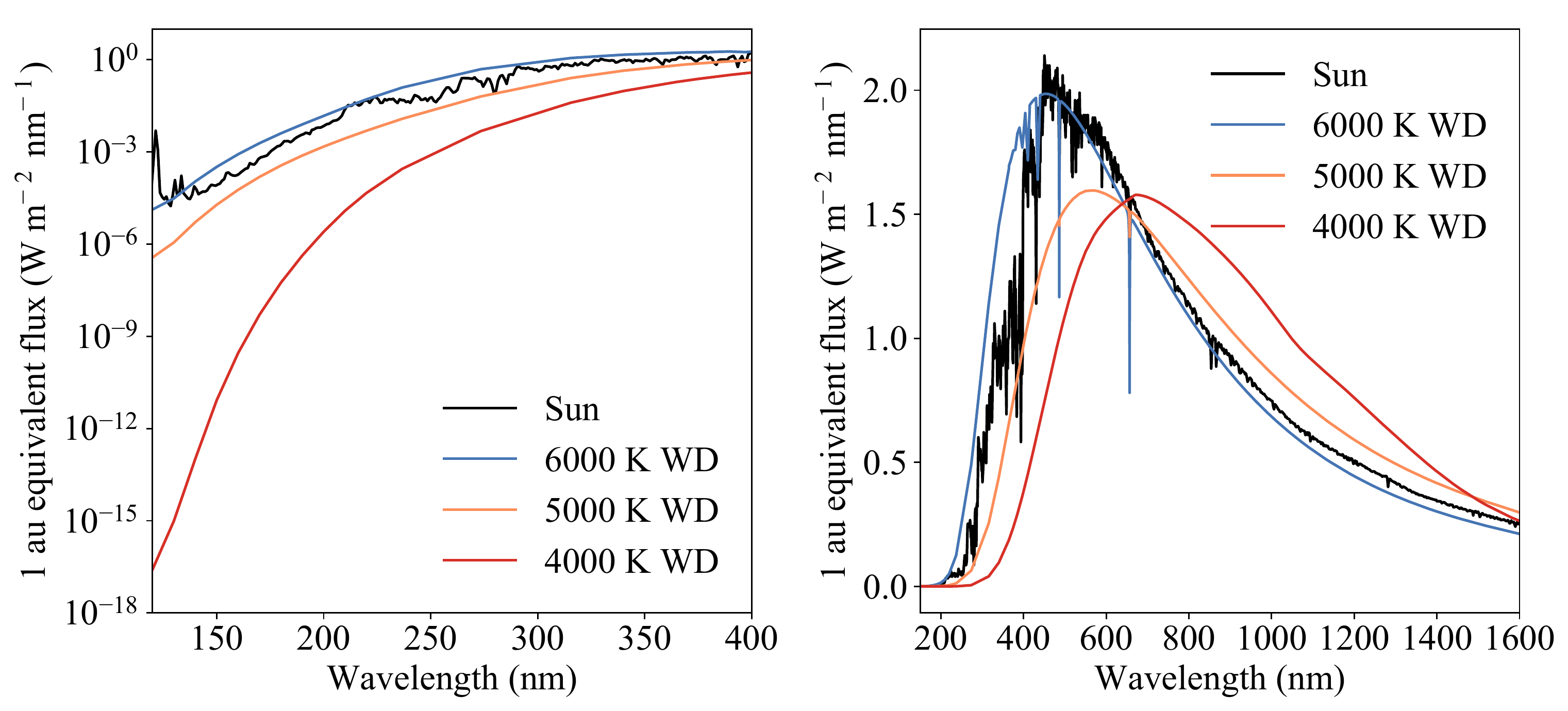}
\caption{Irradiation received at the top of atmosphere for planets orbiting WDs at 1 au equivalent distance. In our models, 1 au equivalent distance means the WD planets receive the same integrated flux as modern Earth around the Sun. The stellar spectra are shown for (left) UV wavelengths and (right) 150-1600 nm. The solar spectrum is shown as black solid lines. The spectra for WDs with effective temperatures of 6000 K, 5000 K, and 4000 K are shown as blue, orange, and red solid lines, respectively. The UV spectra (left) show that 6000 K WD has similar EUV ($< 124$ nm) intensity as the Sun, while 5000 K WD and 4000 K WD have $\sim 10^3$ and $> 10^{10}$ times lower EUV intensity than the Sun, respectively \protect\citep{Saumon_2014ApJ...790...50S}. The overall spectra (right) shows that WD spectra are almost perfect blackbodies while the solar spectrum shows abundant absorption lines, which explains the difficulty of constraining masses of WD exoplanets using the radial velocity method \protect\citep{vanderburg_giant_2020}.  \label{fig:stellar_spectra}}
\end{figure*}
%%% STELLAR SPECTRA ================

While Earth-like composition is an important possibility to include, the atmosphere of an Earth-mass rocky planet may span different oxidation states, from reducing to oxidizing. Therefore, here we attempt to expand our knowledge of putative Earth-mass exoplanets in WD systems by exploring the vast uncharted parameter space from an atmospheric modeling perspective. We follow the ``exoplanet benchmark cases'' outlined by \cite{hu_photochemistry_2012} and consider three types of atmospheric compositions, including reducing (H$_2$-dominated), weakly oxidizing (N$_2$-dominated), and oxidizing (CO$_2$-dominated) atmospheres. Furthermore, the evolutionary history of a first-generation WD exoplanet is distinct from a MS planet like Earth, and the formation origin of a second-generation WD exoplanet is also different from Earth's. Affected by its origin and evolution, WD exoplanets can have atmospheric composition and surface emissions distinct from Earth. Therefore, it is necessary to incorporate the unique evolutionary history of WD exoplanets when modeling their atmospheres.

Due to the large scale height of H$_2$-dominated atmospheres, H$_2$ atmospheres on Earth-mass planets are much easier for JWST to detect and characterize compared to high mean molecular weight (MMW) atmospheres, so we place special emphasis on H$_2$-dominated atmospheres. In addition, for H$_2$ atmospheres, low ultraviolet (UV) radiation environment can facilitate biosignature gases accumulation \citep[e.g.,][]{seager_biosignature_2013}, and we discuss the photochemical implications of H$_2$ atmospheres around cool WDs, which have extremely low UV.
% in combination with low ultraviolet (UV) radiation facilitate biosignature gases accumulation \citep[e.g.,][]{seager_biosignature_2013}, and we discuss the photochemical implications of H$_2$ atmospheres around cool WDs, which have extremely low UV.

% The scale height $H$ is the e-folding factor of the exponential decrease of atmospheric pressure as a function of altitude, and $H$ is defined as $H = kT / \mu m_H g$, where $k$ is the Boltzmann’s constant, T is the atmospheric temperature, $\mu$ is the MMW, $m_H$ is the atomic mass unit, and $g$ is the surface gravity. 
% Due to much smaller MMW, an H$_2$-dominated atmosphere is roughly 14 times more extended than a N$_2$-dominated atmosphere, and roughly 22 times more extended than a CO$_2$-dominated atmosphere.
% The large extent of H$_2$-dominated atmospheres combined with the large planet-to-star size ratio of an Earth-mass planet transiting a WD implies that JWST can detect and characterize an H$_2$-dominated atmosphere with a small number of transits. Given the promising observational prospect, a careful investigation into the evolutionary history of H$_2$-dominated atmospheres on WD planets is necessary.

In what follows, we introduce our input WD stellar spectra and models in Section \ref{sec:methods}. We present our main results for H$_2$-dominated atmospheres in Section \ref{sec:results_h2} and main results for N$_2$- and CO$_2$-dominated atmospheres in Section \ref{sec:results_n2co2}.
% The same methodology and derivation are applied to N$_2$- and CO$_2$-dominated atmospheres, and we discuss results for these two high MMW atmospheres in Section \ref{sec:results_n2co2}. 
Section \ref{sec:discussion} contains the discussion, and we summarize our conclusions in Section \ref{sec:conclusion}.

% \noindent Note that in the two column style figures and tables will only
% span one column unless specifically ordered across both with the ``*'' flag,
% e.g. \\

% \noindent{\tt\string\begin\{figure*\}} ... {\tt\string\end\{figure*\}}, \\
% \noindent{\tt\string\begin\{table*\}} ... {\tt\string\end\{table*\}}, and \\
% \noindent{\tt\string\begin\{deluxetable*\}} ... {\tt\string\end\{deluxetable*\}}. \\

% \noindent This option is ignored in the onecolumn style.

%%======================================================================================
%%  METHODS
%%======================================================================================

\section{Methods} \label{sec:methods}
% In this section, we introduce our models and discuss underlying model assumptions. We present our input WD stellar spectra in Section \ref{sec:stellar_model}. We introduce our photochemical model and the different atmospheric scenarios in Section \ref{sec:photochem_model}. Section \ref{sec:trans_spectra_model} discusses the transmission spectra model and Section \ref{sec:jwst_model} summarizes how we simulate JWST transmission spectra observations for Earth-mass WD exoplanets.

\subsection{Stellar Model} \label{sec:stellar_model}

Following \cite{kozakis_uv_2018, kozakis_high-resolution_2020} we use cool WD spectral models calculated by \cite{Saumon_2014ApJ...790...50S} for WD temperatures of 6000, 5000, and 4000 K to represent WD cooling throughout time. These models assume the average WD mass of 0.6 M$_\odot$ \citep{kepler_new_2016} with pure hydrogen atmospheres and surface gravities of $\log\,g = 8.0$.  Due to the high surface gravity these WD atmospheres are highly differentiated and only display Balmer absorption lines for temperatures $\gtrsim$ 5000 K, with hydrogen becoming neutral at lower temperatures (as seen in Figure \ref{fig:stellar_spectra}).  For further discussions on spectral modeling methods for cool WD atmospheres see e.g., \cite{Saumon_2014ApJ...790...50S}, \cite{kozakis_uv_2018}, and references therein.
% \cite{Bergeron_1997ApJS..108..339B}, \cite{Kowalski_2006ApJ...651L.137K}, \cite{Kilic_2009ApJ...696.2094K, Kilic_2009AJ....138..102K}, \cite{Giammichele_2012ApJS..199...29G}, and \cite{Saumon_2014ApJ...790...50S}.

% WDs with effective temperatures of 6000, 5000, and 4000 K provide stable irradiation environments for planets orbiting them. A WD is initially very hot ($\gtrsim$ 30,000 K) and cools down first quasi-exponentially to $\sim$ 10,000 K \citep[e.g.,][]{fontaine_potential_2001}. While the initial cooling of a hot WD is too rapid to provide a stable irradiation environment for geological timescales, cooling below 6000 K is slow and steady. According to observation based WD cooling models, a typical 0.6 $M_\odot$ WD takes $\sim$ 2 Gyr to cool down to 6000 K, $\sim$ 6 Gyr to 5000 K, and $\sim$ 9 Gyr to 4000 K \citep{Bergeron_2001ApJS..133..413B, fontaine_potential_2001}. Such slow cooling allows a planet to stay in the habitable zone (HZ) for $\sim4$ to $\sim8.5$ Gyr, depending on the planet's orbital distance \citep{kozakis_uv_2018}.

%%% MIXING RATIO ================
\begin{figure*}[!p]
\includegraphics[width=\textwidth]{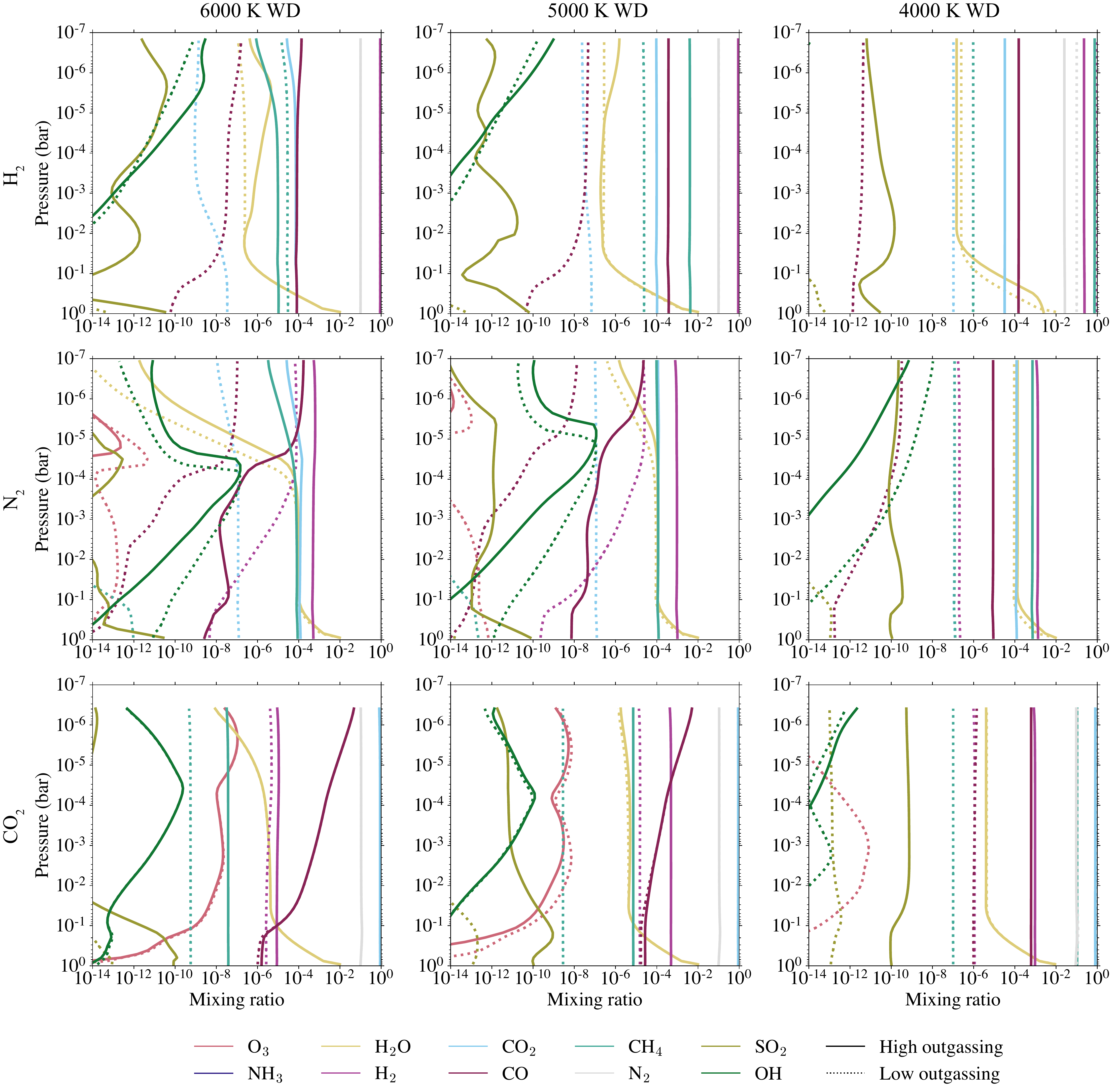}
\caption{Chemical mixing ratios for our model planets. We compare high outgassing rates (solid lines) with low outgassing rates (dotted lines). H$_2$-, N$_2$-, and CO$_2$-dominated atmospheres are shown in the top, middle, and bottom rows, respectively. The left, middle, and right columns show WDs with effective temperatures of 6000 K, 5000 K, and 4000 K, respectively. Mixing ratios of volcanically produced gases, such as CH$_4$, CO$_2$, and SO$_2$, are generally higher in the high outgassing models. Mixing ratios of photochemically destroyed gases such as CH$_4$ are generally higher in the cooler WD models, due to less UV irradiation.  \label{fig:mixing_ratios}}
\end{figure*}
%%% MIXING RATIO ================

\subsection{Photochemistry Model} \label{sec:photochem_model}
We use a one-dimensional photochemistry model that was validated by simulating the atmospheric compositions of modern Earth and Mars \citep{hu_photochemistry_2012}, with CO$_2$ and H$_2$O cross-sections updated by \cite{Ranjan_2020ApJ...896..148R}. The photochemistry model was manually coupled with an analytical atmospheric temperature model formulated by \cite{guillot_radiative_2010} implemented as a part of the petitRADTRANS radiative transfer package \citep{Molliere_2019A&A...627A..67M}. Converged atmospheric chemical profiles are shown in Figure \ref{fig:mixing_ratios}.

We assume three types of anoxic atmospheres following the exoplanet benchmark scenarios presented by \cite{hu_photochemistry_2012}. The benchmark scenarios include a reducing (90\% H$_2$, 10\% N$_2$), a weakly oxidizing ($>99$\% N$_2$), and an oxidizing (90\% CO$_2$, 10\% N$_2$) atmosphere. For each atmospheric scenario, we assume two sets of outgassing rates for CO$_2$, H$_2$, SO$_2$, CH$_4$, and H$_2$S, one corresponding to modern Earth-like volcanic emission rates, and the other corresponding to a less geologically active planet outgassing at 1000 times lower rates. 

We assume Earth-like parameters -- 1 $M_\oplus$ mass, 1 $R_\oplus$ radius, and 1 bar surface pressure -- for all planets modeled. For the N$_2$-dominated atmosphere models, we assume the planet is located at the 1 au equivalent distance, which means that the planet receives the same integrated flux as Earth's irradiation on top of its atmosphere. Due to enhanced greenhouse effect, planets with CO$_2$- and H$_2$-dominated atmospheres are placed at 1.3 au and 1.6 au equivalent distance to maintain similar surface temperature conditions. For more details of the photochemistry model, refer to Appendix \ref{sec:ap_chem_details}.

\subsection{Transmission Spectra Model} \label{sec:trans_spectra_model}
We use a publicly available Python package petitRADTRANS \citep{Molliere_2019A&A...627A..67M} to calculate the transmission spectra for all our planet models, using our photochemistry code results as inputs. We apply the low-resolution correlated-k method to generate spectra with $\lambda / \Delta \lambda = 1000$. We generate 100 atmospheric layers with pressures distributed equidistantly in log space, starting from the surface (1 bar) to the top of atmosphere ($\sim 10^{-7}$ bar).

Cloud layers mute spectral signatures by blocking molecules below from the view of an observer (see e.g., \citealt{Komacek_2020ApJ...888L..20K, Suissa_2020ApJ...891...58S} for recent 3D explorations). In addition, atmospheric refraction may prevent transmission spectroscopy from accessing the lower atmosphere by bending light rays away from a distant observer \citep[e.g.,][]{betremieux_2014ApJ...791....7B, Robinson_2017ApJ...850..128R}. Effects of clouds and refraction are both accounted for in our transmission spectra model. For more details of the transmission spectra model, refer to Appendix \ref{sec:ap_tran_details}.

%%% FLOWCHART ================
\begin{figure*}[ht!]
\centering
\includegraphics[width=0.9\textwidth]{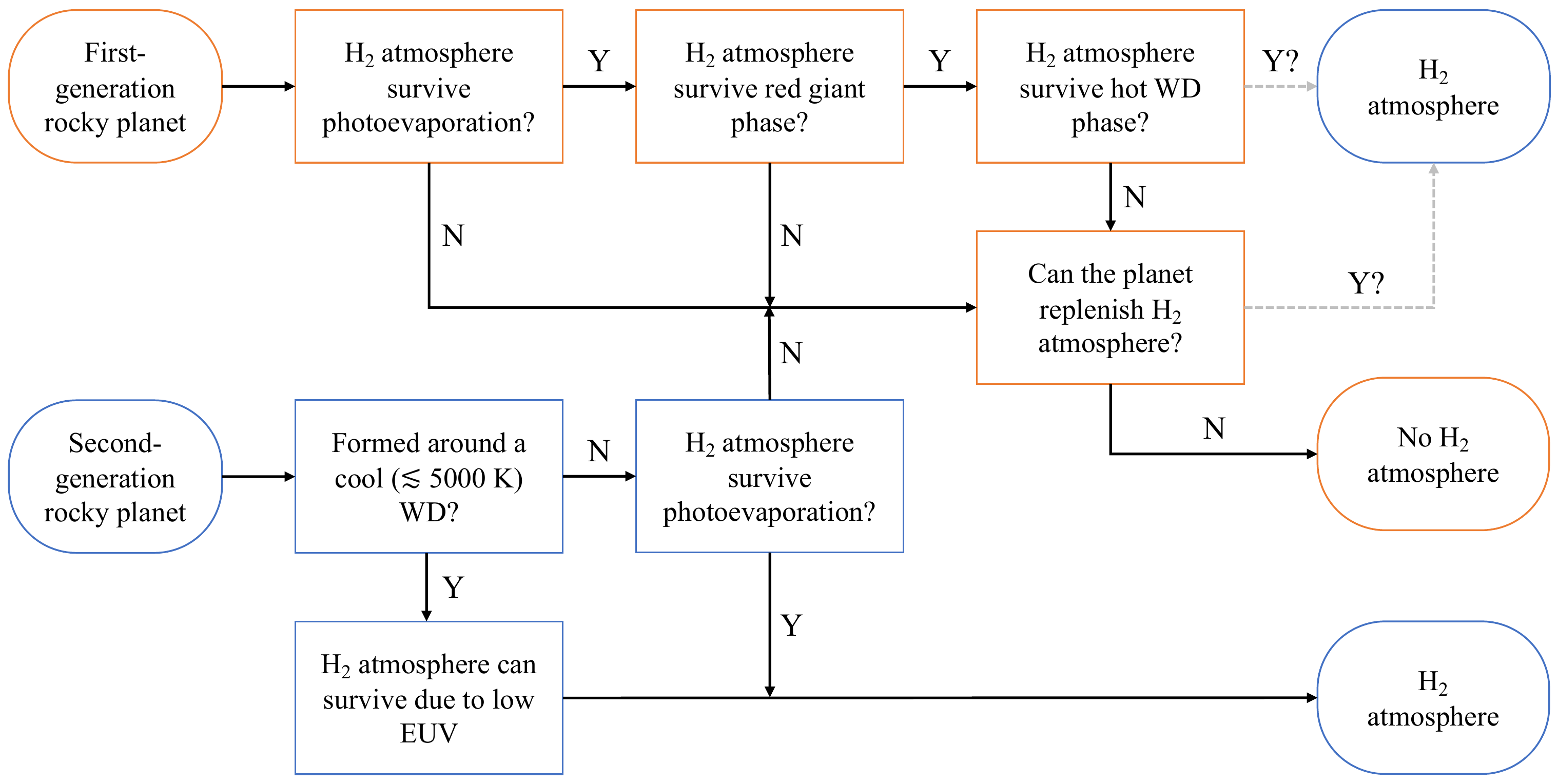}
\caption{Flowchart showing evolution of H$_2$ atmospheres on first- and second-generation rocky WD exoplanets. Plausible evolutionary pathways are shown as solid black arrows, while improbable pathways are shown as dashed grey arrows. The main takeaway from this flowchart is that retention of H$_2$ atmospheres on first-generation Earth-mass rocky planets requires a sequence of unlikely coincidences, while second-generation rocky WD planets have a viable pathway to produce and maintain a H$_2$-dominated atmosphere.  \label{fig:flowchart}}
\end{figure*}
%%% FLOWCHART ================

\subsection{Simulated JWST Observations} \label{sec:jwst_model}
We simulate JWST observations of our model rocky WD exoplanets with the NIRSpec Prism instrument using PandExo \citep{batalha_pandexo_2017}. 
% We consider two JWST instruments, NIRSpec Prism and MIRI LRS, covering wavelength ranges 0.6--5.3 $\micron$ and 5--14 $\micron$, respectively. We find that MIRI LRS does not provide new spectral features while having much lower signal-to-noise ratio (SNR) compared to NIRSpec Prism. Therefore, we focus on NIRSpec Prism.
For all simulations, we limit saturation level to 80\% and do not bin the output spectra. For stellar input parameters, we use the physical properties of WD 1856+534 reported by \cite{vanderburg_giant_2020}: J band magnitude = 15.677 and stellar radius = 0.0131 $R_\odot$. For planet parameters, we use the transmission spectra generated by petitRADTRANS as the input spectra, assuming a 1 $R_\oplus$ planet radius and a 2-minute transit duration, which is the transit duration of a planet in the HZ of WD 1856+534 (e.g., \linktocite{kaltenegger_white_2020}{Kaltenegger, MacDonald et al.\ 2020}). All other observational parameters are optimized by PandExo under the default settings. We simulated a range of JWST campaign sizes: 1 transit, 5 transits, 10 transits, and 25 transits, which corresponds to 2, 10, 20, and 50 minutes of total in-transit integration time, respectively.

%%======================================================================================
%%  RESULTS - H2
%%======================================================================================

\section{Results} \label{sec:results_h2}
Here we demonstrate in Section \ref{sec:31_h2_indicator} that a thick ($\sim 1$ bar) H$_2$-dominated atmosphere on a rocky WD exoplanet is an indicator of a second-generation Earth-mass planet. We show that such H$_2$ atmospheres are easily detectable by JWST and discuss the implications of our simulated transmission spectra in Section \ref{sec:32_spectra_diff}--\ref{sec:34_h2_test_runaway}. Detectability of key spectral features by JWST is summarized in Appendix \ref{sec:ap_h2_features}.

\subsection{H$_2$-dominated Atmospheres as Indicators of Second-generation Planets} \label{sec:31_h2_indicator}
Here we we qualitatively discuss the evolution of a H$_2$-dominated atmosphere. Quantitative results supporting the qualitatively analysis are shown in Appendix \ref{sec:ap_quantitative}.

\subsubsection{Evolution of H$_2$-dominated Atmospheres on WD Exoplanets} \label{sec:311_evolution}
% In this subsection, we discuss how an H$_2$-dominated atmosphere evolves under the context of the evolutionary history of a WD system. We review the two distinct evolutionary paths tracked by first- and second-generation Earth-mass planets and show that as a natural result of the two paths, an H$_2$-dominated atmosphere can only be detected on second-generation planets. Evolutionary history of both first- and second-generation rocky WD planets are summarized in Figure \ref{fig:flowchart}.

We begin our evolutionary analysis by considering an Earth-mass planet formed around the young MS progenitor (top row, Figure \ref{fig:flowchart}). Primary H$_2$ envelopes of an Earth-mass rocky planet may experience photoevaporation due to the excessive EUV radiation produced by young stars \citep[e.g.,][]{owen_kepler_2013, Lopez_2013ApJ...776....2L}. A terrestrial planet either loses its hydrogen atmosphere entirely and remains as a barren core with a radius distribution that peaks at $\sim 1.3\,R_\oplus$, or retains a very thick envelope that doubles the core’s radius, creating an ``evaporation valley'' in the radius distribution of known exoplanets \citep[e.g.,][]{Owen_2017ApJ...847...29O}. A more recent work argues that $\sim 2\,M_\oplus$ rocky planets with H$_2$ envelopes may have already been discovered, and even lower mass planets with voluminous H$_2$ atmospheres possibly exist \citep{owen_hydrogen_2020}. Hence, first-generation H$_2$-rich rocky planets cannot be ruled out on the basis of photoevaporation. 

Even if the H$_2$ envelope on an Earth-mass exoplanet survives the MS phase, it is likely to be lost in the post-MS evolution. The red giant phase and the hot WD phase can both threaten H$_2$ atmospheres.
% When a star turns into a red giant after the end of its MS lifetime, the gaseous envelope of the star can expand to a fraction of an au or even a few au depending on stellar mass, eroding planetary atmospheres with stellar winds, and even engulfing close-in planets (see e.g., \citealt{Bertelli_2008A&A...484..815B, Bertelli_2009A&A...508..355B} for stellar evolution tracks). 
As a red giant loses its gaseous envelope, stellar winds due to mass loss can erode atmospheres \citep[e.g.,][]{ramirez_habitable_2016, kozakis_atmospheres_2019}. 
% A $1\,M_\oplus$ planet with a 1 bar high MMW atmosphere at a distance that receives Mars equivalent stellar radiation would lose 4\% of its atmosphere when its $1.9\,M_\odot$ (A5 type on MS) host turns into a red giant. If mass of the host star is $1.3\,M_\odot$ or smaller (F5 or later types on MS), atmospheric loss fraction rises to $\sim 100\%$ at the same Mars equivalent distance \citep{ramirez_habitable_2016}. 
Outward migration is therefore necessary for atmospheric retention. Such orbital expansion can be triggered by red giant mass loss \citep[e.g.,][]{Schroder_2008MNRAS.386..155S}.

\begin{figure*}[ht]
\includegraphics[width=\textwidth]{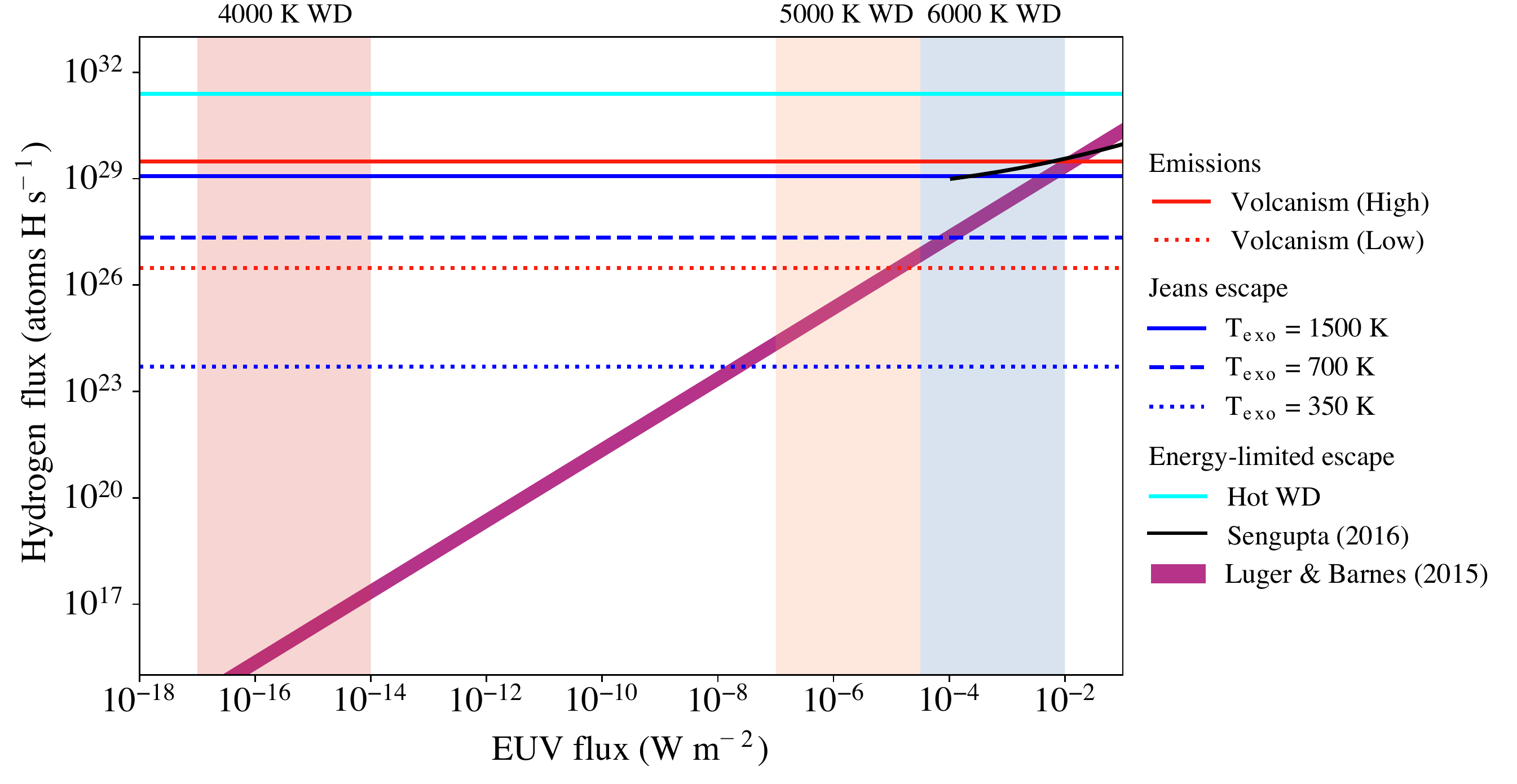}
\caption{Volcanic emission rates and escape rates of hydrogen in atmospheres of $1\,M_\oplus$ rocky WD exoplanets. The energy-limited escape rates (purple and black solid lines) are functions of EUV flux received by the planet, while all other lines are independent of flux or assume a fixed flux. The red solid line and the red dotted line represent high and low volcanic emission rates of H$_2$, respectively. The high emission rate is 20 times higher than modern Earth H$_2$ emission rate and 1000 times higher than the low emission rate. The blue solid line, blue dashed line, and blue dotted line demarcate the Jeans escape rates assuming exobase temperatures of 1500 K (roughly the maximum exospheric temperature on Earth), 700 K (roughly the minimum exospheric temperature on Earth), and 350 K (exobase temperature on Mars), respectively. The cyan solid line is the energy-limited escape rate of a planet orbiting a 40,000 K hot WD at a distance of $\sim 50$--$80$ au \protect\citep{Schreiber_2019ApJ...887L...4S}. The black solid line is the energy-limited escape rate calculated based on the \protect\cite{watson_dynamics_1981} and \protect\cite{sengupta_upper_2016} formulas, which are only solvable for a certain range. 
% Note that the \cite{sengupta_upper_2016} equations are only solvable for a certain range of H escape fluxes and terminates below $\sim 10^{29}$ H s$^{-1}$. 
The purple solid line shows a range of energy-limited escape rates assuming different $\epsilon_{EUV}$ and $R_{EUV}$ for a sensitivity test \protect\citep{luger_extreme_2015}. The approximate ranges of EUV radiation received by planets at 1 au equivalent distance orbiting 4000 K, 5000 K, and 6000 K WDs are shown as red, orange, and blue rectangles, respectively. Implications of this figure are: (i) energy-limited escape driven by excessive EUV of hot WDs will erode H$_2$ atmospheres even if the planets have high H$_2$ emission rate, and (ii) maintaining H$_2$-dominated atmospheres is possible for planets with high outgassing rates orbiting all three types of WDs, and is possible for planets with low outgassing rates orbiting cool ($\lesssim 5000$ K) WDs, assuming suitable exobase temperatures. For numerical details, see Appendix \ref{sec:ap_quantitative}. \label{fig:H_emissions_escape}}
\end{figure*}

A hydrogen-dominated atmosphere on an Earth-mass planet is very unlikely to survive the hot WD phase, even if it fortuitously survived both the MS phase and the red giant phase. Young WDs initially have very high effective temperatures ($T_{\rm eff}$) and intense EUV radiation. When a solar mass star turns into a WD, it will start with $T_{\rm eff} \gtrsim 100,000$ K and then experience quasi-exponential radiative cooling to $\sim30,000$ K in a timespan of approximately 10 Myr \citep{fontaine_potential_2001}. High $T_{\rm eff}$ results in extreme EUV intensity. An exoplanet orbiting a hot WD will be bombarded by excessive EUV radiation up to a million times higher than modern solar levels, resulting in rapid atmospheric mass loss even if the planet migrates to an orbital distance of 50--100 au \citep{Schreiber_2019ApJ...887L...4S}. Volcanic H$_2$ emission cannot compensate such loss (Figure \ref{fig:H_emissions_escape}). We therefore conclude that the EUV radiation from hot WDs will cause total erosion of hydrogen atmospheres. In Appendix \ref{sec:ap_quantitative} we quantitatively prove this conclusion.

Given that the survival of a primary hydrogen atmosphere on an Earth-mass first-generation WD exoplanet is highly unlikely, replenishing a H$_2$-dominated atmosphere via outgassing remains as the only possibility. We will show in Section \ref{sec:312_first_gen} that replenishing hydrogen by outgassing is unlikely for a first-generation planet.

%%% H ESCAPE AND EMISSIONS ================
%%% H ESCAPE AND EMISSIONS ================

We now shift the focus to second-generation Earth-mass planets and show that planets formed in the debris disks of WDs can possibly maintain a detectable H$_2$-dominated atmosphere (bottom row, Figure \ref{fig:flowchart}). Protoplanets can form in tight orbits around WDs via coagulation of viscously spreading disk materials, and then the protoplanets can further accrete disk material to form major planets just outside the Roche limit of the host WD. If a super-Earth is tidally destroyed, this formation mechanism can potentially recycle materials, including volatile materials such as water, from the disrupted planet to form an Earth-mass planet \citep{bear_planetary_2015, vanlieshout_exoplanet_2018}.

Second-generation planets can form at any time during a WD’s lifetime. Formation of second-generation WD planets is possible whenever planetary materials are delivered into a WD’s Roche limit and be tidally disrupted \citep{bear_planetary_2015, vanlieshout_exoplanet_2018}. Such delivery can occur at any time during a WD’s lifetime, because WD pollutants have been detected on WDs across temperature range. The coolest polluted WD has an effective temperature of $\sim 4000$ K \citep{Coutu_2019ApJ...885...74C}, evidencing that some of the oldest known WDs are actively accrete materials, so second-generation planet formation can occur around cool WDs.

%%% TRANSMISSION SPECTRA W/ PANDEXO, H2 ================
\begin{figure*}[!p]
\centering
\includegraphics[width=0.7\textwidth]{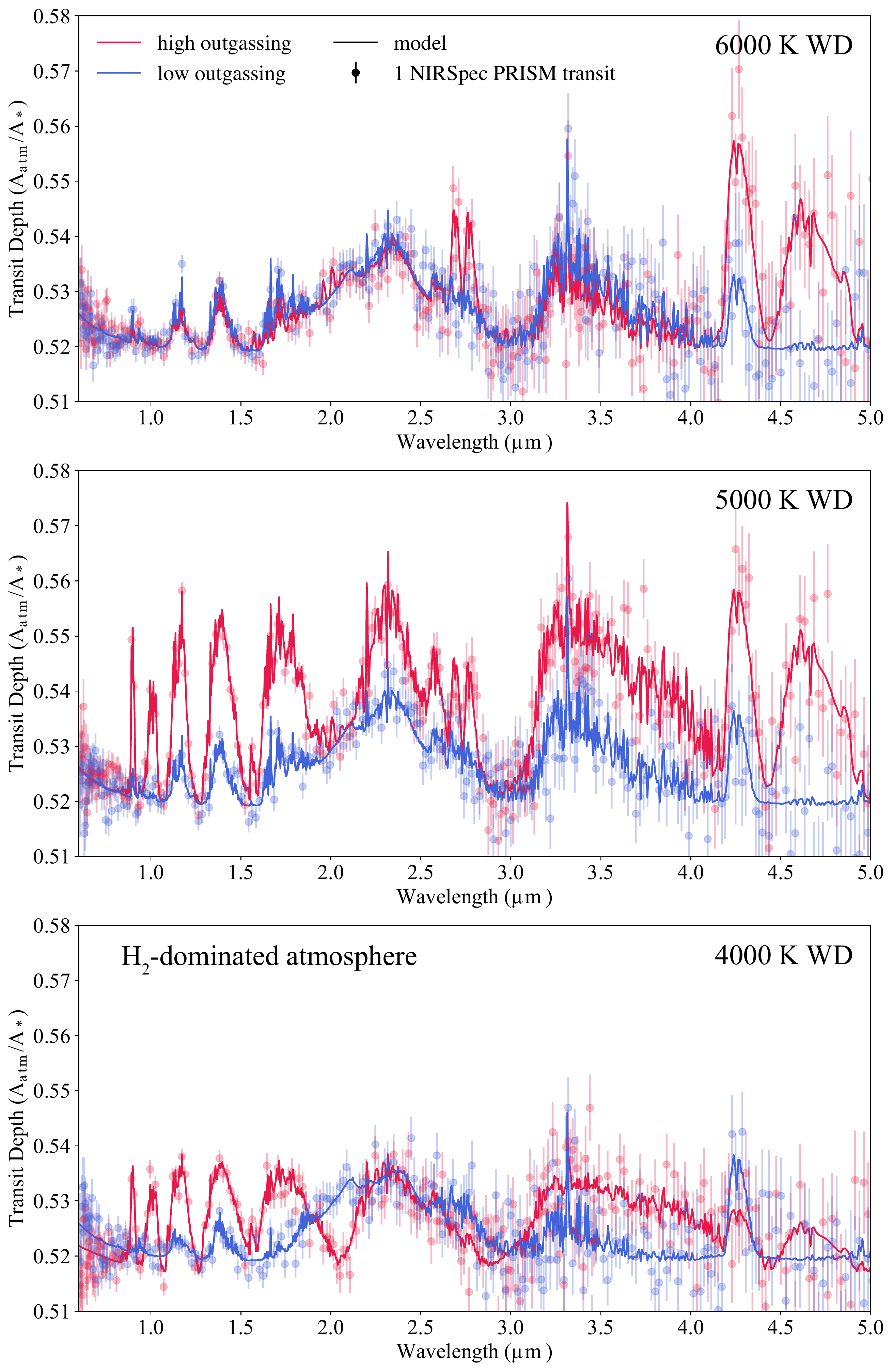}
\caption{Transmission spectra models and simulated JWST transit observations for H$_2$-dominated atmospheres, assuming a $1\,R_\oplus$ planet transiting WD 1856+534. We compare high outgassing rate models (red solid lines) with low outgassing rate models (blue solid lines) and show simulated JWST data points with their $1\,\sigma$ error bars. We assume 1 NIRSpec Prism transit for the H$_2$-dominated atmosphere models. In the H$_2$-dominated atmospheres, spectral signatures have high detectability, and differentiating between high and low outgassing scenarios is achievable with 1 transit.  \label{fig:spectra_pandexo_h2}}
\end{figure*}
% H$_2$-, N$_2$-, and CO$_2$-dominated atmospheres are shown in the top, middle, and bottom rows, respectively. The left, middle, and right columns show WDs with effective temperatures of 6000 K, 5000 K, and 4000 K, respectively. Note that we assume 1 NIRSpec Prism transit for the H$_2$-dominated atmosphere models and 25 NIRSpec Prism transits for the N$_2$-, and CO$_2$-dominated models. In the H$_2$-dominated atmospheres, spectral signatures have high detectability, and differentiating between high and low outgassing scenarios is achievable with 1 transit. With 25 transits, major spectral features are also detectable in the high MMW atmosphere models, but whether the high and low outgassing scenarios are distinguishable depends on whether CH$_4$ accumulates to much higher level in the high outgassing models.
%%% TRANSMISSION SPECTRA W/ PANDEXO, H2 ================

Second-generation planets formed around $> 6000$ K WDs are also vulnerable to photoevaporation because of their modern Sun-like UV radiation (Figure \ref{fig:stellar_spectra}).

A hydrogen atmosphere can survive, however, if the planet forms around a cool ($\lesssim 5000$ K) WD. EUV intensity of a 5000 K WD is $\sim 100$--$1000$ times less than modern Sun, while a 4000 K WD emits $\gtrsim 10^{10}$ times less EUV radiation than modern Sun (Figure \ref{fig:stellar_spectra}). Both types of dominant neutral atmospheric escape mechanisms -- Jeans escape \citep[e.g.,][]{Hunten_1973JAtS...30.1481H} and hydrodynamic escape \citep{watson_dynamics_1981, luger_extreme_2015, sengupta_upper_2016} -- are powered by EUV radiation. We will show quantitatively in Appendix \ref{sec:ap_quantitative} that given the low EUV intensity of $\lesssim 5000$ K WDs, high to moderate volcanic hydrogen emission rates are sufficient to sustain a H$_2$-dominate atmosphere (Figure \ref{fig:H_emissions_escape}).

In summary, the most viable path to a roughly 1 bar H$_2$-dominated atmosphere on an Earth-mass WD exoplanet is provided by second-generation planets formed around cool WDs (Figure \ref{fig:flowchart}). Survival of H$_2$ atmospheres on first-generation rocky WD exoplanets requires a sequence of improbable coincidence.

\subsubsection{First-generation Planets Cannot Replenish H$_2$ Atmosphere via Outgassing} \label{sec:312_first_gen}
Once the H$_2$-dominated atmosphere on a first-generation rocky planet is lost, replenishing a secondary H$_2$ atmosphere via outgassing is unlikely. Replenishing a reduced hydrogen atmosphere requires reducing volcanic emissions, neglecting other sources such as biogenic emission. Reduced volcanic emissions require reduced mantles. The dominant form of hydrogen outgassing on a planet with an oxidized mantle, such as present-day Earth, is H$_2$O \citep{Holland_1984ceao.book.....H}. Molecular hydrogen is the dominant form of hydrogen emission only when the planet’s mantle is highly reduced \citep{ramirez_warming_2014, Ortenzi_2020NatSR..1010907O}. 
% Even with high H$_2$ outgassing flux, when other oxidizing gases are emitted simultaneously under a weakly or strongly oxidizing environment, H$_2$ cannot accumulate to a substantial fraction, as evidenced by photochemical models for our N$_2$- and CO$_2$-dominated atmospheres (Figure \ref{fig:mixing_ratios}).

%%% TABLE 2 - RULING OUT A FLAT LINE ================
\begin{table*}[!bt]
\caption{$\sigma$ significance of ruling out a flat line based on transmission spectra}
\label{table:sigma_flat_line}

\centering
\begin{tabular}{lccccccccc}
\hline\hline
\multicolumn{10}{l}{\textit{High outgassing}} \\
 & \multicolumn{3}{c}{N$_2$ atmospheres}   & \multicolumn{3}{c}{CO$_2$ atmospheres}     & \multicolumn{3}{c}{H$_2$ atmospheres}     \\
 & 4000 K & 5000 K & 6000 K & 4000 K & 5000 K & 6000 K & 4000 K & 5000 K & 6000 K  \\
\hline

1 NIRSpec Prism transit  & 2.7 & 5.2 & 6.6 & 5.1 & 4.7 & 2.3 & $>10$ & $>10$ & $>10$ \\
5 NIRSpec Prism transits & $>10$ & 6.8 & 7.5 & 7.5 & 7.8 & 7.4 & $>10$ & $>10$ & $>10$ \\
\hline

\multicolumn{10}{l}{\textit{Low outgassing}}   \\
 & \multicolumn{3}{c}{N$_2$ atmospheres}   & \multicolumn{3}{c}{CO$_2$ atmospheres}     & \multicolumn{3}{c}{H$_2$ atmospheres}     \\
 & 4000 K & 5000 K & 6000 K & 4000 K & 5000 K & 6000 K & 4000 K & 5000 K & 6000 K  \\
\hline
1 NIRSpec Prism transit  & 3.4 & 3.7 & 3.6 & 4.3 & 3.7 & 4.4 & $>10$ & $>10$ & $>10$ \\
5 NIRSpec Prism transits & 6.5 & 2.8 & 6.6 & 4.2 & 7.1 & 6.6 & $>10$ & $>10$ & $>10$ \\
\hline

\end{tabular}

\end{table*}
%%% TABLE 2 - RULING OUT A FLAT LINE ================

However, first-generation Earth-mass WD exoplanets are generally expected to have oxidized mantles because of mantle self-oxidation. The mantle of a $\sim 1\,M_\oplus$ rocky planet starts reduced, but becomes more oxidized overtime due to gradual or stepwise self-oxidation \citep{scaillet_redox_2011}. Such self-oxidation is thought to be a natural consequence of the size of Earth-mass rocky planets \citep{wood_accretion_2006}. Similarity in bulk composition between WD exoplanets and the Earth is inferred by WD pollution \citep[e.g.,][]{Farihi_2016MNRAS.463.3186F, doyle_oxygen_2019}. Due to compositional resemblance, we expect Earth-mass first-generation WD planets would undergo the same self-oxidation process.

Mantle self-oxidation of Earth-mass rocky planets occurs within $\sim 100$ Myr \citep{trail_oxidation_2011}, which is much shorter compared to the age of first-generation WD exoplanets. The age of a first-generation WD exoplanet is the sum of MS progenitor lifetime and WD cooling time, which are both on the order of Gyr. Therefore, first-generation rocky WD planets do not have the right conditions to outgas hydrogen in the form of H$_2$.

\subsection{Transmission Spectra Can Differentiate Between First- and Second-generation Planets} \label{sec:32_spectra_diff}

We have demonstrated that theoretically, the presence of H$_2$-dominated atmospheres on Earth-mass WD exoplanets indicates second-generation planets. Observationally, JWST can potentially differentiate between first- and second-generation Earth-mass planets using H$_2$ atmospheres using only 1 NIRSpec Prism transit, assuming a $1\,R_\oplus$ planet transiting WD 1856+534.

An atmosphere is detectable if a null assumption (flat spectrum) can be ruled out conclusively. For each simulated NIRSpec Prism transmission observation, we find the best-fit horizontal line representing a flat spectrum and determine the $\sigma$ significance of ruling out this flat line fit using a $\chi^2$ analysis. Results are summarized in Table \ref{table:sigma_flat_line} and model transmission spectra overplotted with simulated observation data are shown in Figure \ref{fig:spectra_pandexo_h2}. Due to the large scale height of hydrogen atmospheres, 1 transit is sufficient to rule out a flat spectrum for all H$_2$-dominated atmosphere models at $> 10\,\sigma$. 
% The high detectability of H$_2$-dominated atmospheres is a natural consequence of their large extent, as evidenced by modeling studies on rocky exoplanets orbiting M dwarfs \citep[e.g.,][]{seager_biosignature_2013, wunderlich_detectability_2021}. WD exoplanets have stronger transit signals due to the planet-to-star size ratios, so the easy detection of hydrogen atmospheres is not surprising.

High-altitude clouds or hazes layers can hide the atmosphere below, hence decreasing the detectability of the atmosphere. Our models assume an opaque global cloud layer at 0.47 bar ($\approx 35$ km in H$_2$-dominated atmospheres). Organic haze formation in atmospheres on rocky Earth-like exoplanets is studied \citep[e.g.,][]{arney_pale_2016, arney_pale_2017, arney_organic_2018} but not included in our atmospheric models, and will be discussed in Section \ref{sec:53_haze}.

\subsection{Transmission Spectra Can Infer Outgassing Activities} \label{sec:ap_h2_outgas}
Rocky WD exoplanets can have drastically different outgassing rates depending on their tectonic states. An Earth-mass rocky planet can possibly have three tectonic states: (i) active lid, where lithosphere strength is overwhelmed by convective stresses and surface materials can be recycled into the mantle, (ii) stagnant lid, where the lithosphere is too rigid to be deformed and recycled into the mantle, and (iii) an episodic regime characterized by occasional lithosphere overturn (see e.g., \citealt{lenardic_climate-tectonic_2016} and references therein).

Conceptually, planets with Earth-like active plate tectonics can recycle atmospheric and oceanic volatiles back into the mantle to sustain continuous outgassing over long timescales \citep[e.g.,][]{kasting_evolution_2003}, while stagnant lid planets cannot. Indeed, some argued that Earth-mass stagnant lid planets would deplete mantle volatiles rapidly and the outgassing rates would drop to negligible levels within $\sim 1$--$2$ Gyr \citep[e.g.,][]{foley_carbon_2018, dorn_outgassing_2018}. 
% WD planets are typically old ($\sim 3$--$10$ Gyr).
% -- the age of WD planets is at least on the same order as the WD cooling age ($\sim 2$-$10$ Gyr), except for recently formed second-generation planets. The MS progenitor lifetime would add another $\sim 3$ Gyr to the age of first-generation WD planets.
% \footnote{The $\sim 3$ Gyr lifetime assumption is not arbitrary but rather represents the lifetime of typical MS progenitors. The mass distribution of over 6000 WDs discovered by the Sloan Digital Sky Survey Data Release 12 peaks at around 0.6 $M_\odot$ \citep{kepler_new_2016}. A 0.6 $M_\odot$ WD corresponds to a F0 type MS progenitor according to WD evolution tracks \citep{veras_post-main-sequence_2016}, which typically has a lifetime of 3 Gyr.}
Therefore, differentiating between active lid and stagnant lid regimes on rocky WD exoplanets is ostensibly possible, if we can differentiate between the different outgassing scenarios with transmission spectroscopy.

However, the topic of rocky planet tectonics remains hotly debated, and the possibility of inferring tectonic activities from outgassing rates is challenged by several uncertainties. One major uncertainty is that rocky planets can alternate between multiple tectonic states over Gyr timescales \citep{weller_effects_2015}, so we are agnostic about a planet’s tectonic history given only a snapshot of its present-day outgassing rates. In addition, even for the best-studied rocky planet -- Earth -- the onset and end of plate tectonics are highly uncertain (see e.g., \citealt{rey_spreading_2014, weller_effects_2015} and references therein).
% Another uncertainty comes from the onset and end of plate tectonics. Even for the best-studied rocky planet -- Earth -- a consensus has not been reached for when and how plate tectonics started (see e.g., \citealt{rey_spreading_2014} and references therein). Earth will ultimately transition to the stagnant lid regime when plate tectonics terminate, but the timing of such transition is also under debate (see e.g., \citealt{weller_effects_2015} and references therein). Finally, we are uncertain about the preferred tectonic states of rocky exoplanets. Different models give different predictions about whether active lid or stagnant lid is preferred by rocky exoplanets, especially rocky super-Earths that are about a few times more massive than Earth \citep[e.g.,][]{ONeill_2007GeoRL..3419204O, Stein_2013E&PSL.361..448S, Valencia_2007ApJ...670L..45V, Tackley_2013Icar..225...50T}, while some argue that it is the evolutionary pathway, rather than compositions or thermal parameters, that determines the tectonic state of a terrestrial planet \citep[e.g.,][]{Lenardic_2012ApJ...755..132L}.

Due to the uncertainties of rocky planet tectonic states, we do not intend to infer the tectonic activities or evolutionary history of rocky WD exoplanets from their atmospheres. Instead, we focus on the capability of transmission spectra to differentiate between high and low outgassing scenarios, and leave the linkage between outgassing rates and tectonic states for future investigation. Here we consider planets with a thick ($\sim 1$ bar) atmosphere and two outgassing rates: a high outgassing rate corresponding to modern Earth-like volcanic emission rates (following \citealt{hu_photochemistry_2012}), and a low outgassing rate that is reduced by a factor of 1000.

Transmission spectra can differentiate between our high and low outgassing scenarios assuming H$_2$-dominated atmospheres. The two outgassing scenarios can be differentiated because strong spectral absorbers such as CH$_4$ and CO$_2$ reach higher equilibrium mixing ratio when surface emission fluxes are higher. Trace gases in H$_2$ atmospheres with low UV irradiation can easily accumulate \citep[e.g.,][]{seager_biosignature_2013}, which further amplifies the difference between the two scenarios.

We demonstrate the difference between our high and low outgassing scenarios and the ability of JWST to differentiate them in Figure \ref{fig:spectra_pandexo_h2}, assuming 1 NIRSpec Prism transit. CH$_4$ accumulates to very high levels in the cool (4000 and 5000 K) WD models for reasons we will discuss in Section \ref{sec:34_h2_test_runaway}. As a result, the high and low outgassing scenarios can be differentiated at $\sim 3\,\sigma $ and $> 5\,\sigma$ in the 4000 K and 5000 K models, respectively, based on several CH$_4$ features in 1.0--2.5 $\micron$. For the 6000 K WD model, CO$_2$ features at 2.8 and 4.2 $\micron$ and the CO feature at 4.7 $\micron$ are the keys differences. Conclusive differentiation is not achievable for the 6000 K WD model with 1 NIRSpec Prism transit due to low singal-to-noise ratio (SNR), but is achievable with 5 transits.

%%% INDIVIDUAL SPECIES H2 ================
\begin{figure*}[t!]
\includegraphics[width=\textwidth]{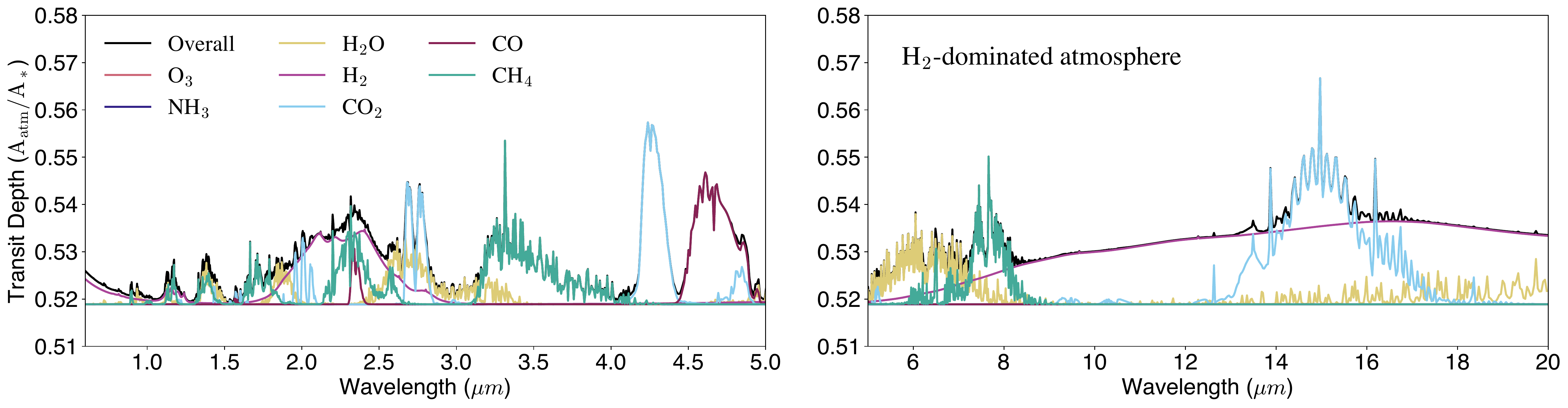}
\caption{Molecular contributions to the overall model transmission spectra for H$_2$-dominated atmospheres. Only the 6000 K WD high outgassing rate scenario is shown for simplicity. The left panel shows features in the bandpass of NIRSpec Prism (0.6-5 $\micron$) to allow comparison with Figure \ref{fig:spectra_pandexo_h2}, and the right panel shows infrared features in 5-20 $\micron$. \label{fig:individual_species_h2}}
\end{figure*}
% Molecular contributions to the overall model transmission spectra for H$_2$-, N$_2$-, and CO$_2$-dominated atmospheres (top to bottom). Only planets with high outgassing rates around 6000 K WDs are shown for simplicity. Only features in the bandpass of NIRSpec Prism (0.6-5 $\micron$) are shown to allow comparison with Figure \ref{fig:spectra_pandexo}.
%%% INDIVIDUAL SPECIES H2 ================

\subsection{H$_2$-Dominated Atmospheres Around WDs Can Test Photochemical Runaway} \label{sec:34_h2_test_runaway}

%%% TABLE 3 - CH4 BUILDUP ================
\begin{table*}[!tb]
\caption{Column-integrated CH$_4$ Mixing Ratios in Photochemistry Models}
\label{table:ch4_buildup}

\centering
\begin{tabular}{lcccccc}
\hline\hline
& \multicolumn{2}{c}{H$_2$ atmospheres} & \multicolumn{2}{c}{CO$_2$ atmospheres} & \multicolumn{2}{c}{N$_2$ atmospheres} \\
& \multicolumn{1}{l}{high outgassing} & \multicolumn{1}{l}{low outgassing} & \multicolumn{1}{l}{high outgassing} & \multicolumn{1}{l}{low outgassing} & \multicolumn{1}{l}{high outgassing} & \multicolumn{1}{l}{low outgassing} \\
\hline

4000 K	& 0.73	& $9.5 \times 10^{-7}$	& 0.1	& $1.0 \times 10^{-7}$	& $7.1 \times 10^{-4}$	& $1.2 \times 10^{-7}$ \\
5000 K 	& $4.2 \times 10^{-3}$	& $2.4 \times 10^{-5}$	& $7.2 \times 10^{-6}$	& $2.8 \times 10^{-9}$	& $1.2 \times 10^{-4}$	& $1.7 \times 10^{-13}$  \\
6000 K	& $1.0 \times 10^{-5}$	& $3.0 \times 10^{-5}$	& $3.8 \times 10^{-8}$	& $5.7 \times 10^{-10}$	& $8.5 \times 10^{-5}$	& $7.4 \times 10^{-13}$ \\
\hline

\end{tabular}

\end{table*}
%%% TABLE 3 - CH4 BUILDUP ================

Gases whose major sinks are photochemical reaction powered by UV photons can potentially undergo so-called ``photochemical runaway'' in H$_2$-dominated atmospheres around cool WDs. Photochemical runaway occurs when the emission of a gas saturates its photochemical sink. Under such conditions, trace species, such as biosignature gases, can accumulate to detectable levels \citep[e.g.,][]{Segura_2005AsBio...5..706S, sousa-silva_phosphine_2020, Zhan_2021AsBio..21..765Z, ranjan_photochemical_2022}. It is debated whether photochemical runaway is a physically realistic process or an artifact of photochemistry models. For a detailed quantitative exploration in support of the physicality of photochemical runaway, see \cite{ranjan_photochemical_2022}.

Photochemical runaway can occur on exoplanets around cool WDs because the UV-powered photochemical sinks can be easily saturated. UV photons generally play a key role in the photochemical removal of atmospheric gases for planets and moons in the Solar System, either by dissociating molecules directly or by producing reactive radicals \citep[e.g.,][]{catling_kasting_2017, ranjan_photochemical_2022}. As the effective temperature of a WD drops from 6000 K to 4000 K, UV radiation plummets by a factor of $\sim 10^{10}$ (Figure \ref{fig:stellar_spectra}).

Extremely low UV radiation from cool WDs implies that a physically plausible surface emission flux can trigger photochemical runaway. Indeed, runaway buildup is observed for CH$_4$, CO$_2$, and CO in our H$_2$-dominated atmosphere models (Figure \ref{fig:mixing_ratios}). Here we use CH$_4$ as an example to study photochemical runaway in H$_2$ atmospheres around cool WDs, because CH$_4$ buildup has the strongest impact on transmission spectra. Rapid CH$_4$ buildup as $T_{\rm eff}$ of the host WD decreases agrees with previous study assuming Earth-like atmospheres \citep{kozakis_uv_2018}. Column integrated mixing ratios of CH$_4$ are summarized in Table \ref{table:ch4_buildup}.

The nonlinear nature of photochemical runaway is observed in our models. Once past the runaway threshold, CH$_4$ increase is thought to increase drastically given a small increase in surface emission \citep{ranjan_photochemical_2022}. Indeed, a 1000-fold increase in outgassing rate results in a $\sim 10^6$ increase in CH$_4$ mixing ratio in our 4000 K WD H$_2$ atmosphere model. For comparison, the same increase only results in a $\sim 100$ times increase in mixing ratio in the 5000 K WD model, which can be explained by higher runaway threshold because 5000 K WD produces higher UV radiation. In the 6000 K WD model, an increase in CH$_4$ outgassing rate results in a slight decrease in CH$_4$ mixing ratio. This is likely because outgassing rates of oxidizing gases such as CO$_2$ are increased by the same factor, and the supply of UV photons is sufficient to catalyze the reaction of CH$_4$ with oxidants. In other words, for $>6000$ K, pCH$_4$ is limited by oxidant flux, not UV flux, as pO$_2$ is limited by reductant flux on modern Earth. Given the modern Sun-like UV radiation of a 6000 K WD, an increase in oxidizing gas emission has a stronger impact on the photochemical equilibrium than an increase in CH$_4$.

Runaway buildup of CH$_4$ on transiting cool WD exoplanets also implies that spectral signatures of CH$_4$ are easily detectable. For H$_2$-dominated models, 1 transit with NIRSpec Prism can detect CH$_4$ conclusively (Figure \ref{fig:spectra_pandexo_h2}) for a hypothetical rocky planet orbiting WD 1856+534 in the WD HZ. For reference, Figure \ref{fig:individual_species_h2} shows the contribution by each important absorbing species to the overall transmission spectra. For the H$_2$ 6000 K scenario, models fit simulated observations very well at the 1.2 and 1.4 $\micron$ CH$_4$ features, with the highest data points $\sim 5\,\sigma$ from the baseline. The CH$_4$ feature at 3.4 $\micron$ is also accessible despite larger error bars. The CH$_4$ feature at 2.3 $\micron$, however, overlaps with a H$_2$-H$_2$ CIA feature and hence is challenging to detect. For the H$_2$ 5000 K scenario, the high CH$_4$ mixing ratio ($4.2 \times 10^{-3}$) combined with the extended scale height of a H$_2$-dominated atmosphere produce very strong CH$_4$ feature for the high outgassing case. The strength of the features make very high $\sigma$ significance detection possible with 1 NIRSpec Prism transit, especially in the shorter wavelengths due to high SNR. For the H$_2$ 4000 K model, the runaway buildup of CH$_4$ that reaches a mixing ratio of 0.73 significantly changes the atmosphere’s mean molecular weight and shrinks its scale height. CH$_4$ features in the high outgassing model are weaker than in the 5000 K scenario, but still strong enough for $> 5\,\sigma$ detection. In the low outgassing model, CH$_4$ features are considerably weaker, but highest data points at the 1.4 $\micron$ feature are still $\sim 3\,\sigma$ from baseline.

Our examination on CH$_4$ is only a case study -- photochemical runaway can occur for any gas with high enough production rate to saturate its photochemical sink, which is usually powered by UV photons. The extremely low UV radiation of cool WDs means that many gases, including CO, PH$_3$, NH$_3$, and isoprene, can easily enter photochemical runaway \citep[e.g.,][]{ranjan_photochemical_2022}. Among these gases, PH$_3$, NH$_3$, and isoprene have been suggested as potential biosignatures \citep{sousa-silva_phosphine_2020, Huang_2021arXiv210712424H, Zhan_2021AsBio..21..765Z}, while CH$_4$ in combination with O$_2$ or O$_3$ is considered as a strong biosignature pair \citep[e.g.,][]{Lederberg_1965Natur.207....9L}. While we only consider volcanic emission, on habitable WD planets, additional biogenic emission may trigger photochemical runaway more easily. Therefore, transiting terrestrial WD exoplanets may be the most accessible targets for biosignature detection.

%%======================================================================================
%%  RESULTS FOR N2 AND CO2 ATMOSPHERES
%%======================================================================================

\section{Results for N$_2$- and CO$_2$-dominated Atmospheres} \label{sec:results_n2co2}
To explore more possible oxidation states of exoplanet atmospheres, we also run our photochemical model assuming N$_2$- and CO$_2$-dominated atmospheres. The key difference between these two high MMW atmospheric compositions and the H$_2$-dominated scenario is that a WD exoplanet with a high MMW atmosphere can either be first- or second-generation. Here we discuss why this degeneracy is unlikely to be resolved. Additional results for high MMW atmospheres, including outgassing activities, photochemical runaway, and location and detectability of transmission spectra features, are summarized in Appendix \ref{sec:add_results_n2co2}.

% \subsection{First- And Second-generation Earth-mass Planets Cannot Be Differentiated for Planets with High MMW Atmospheres} \label{sec:41_diff_high_mmw}
% We have shown in Section \ref{sec:31_h2_indicator} that the presence of a thick ($\gtrsim 1$ bar) H$_2$ atmosphere indicates a second-generation Earth-mass planet. The presence of high MMW atmospheres, such as the N$_2$- and CO$_2$-dominated atmospheres considered here, however, cannot conclusively determine the formation origin of a rocky WD exoplanet.

Ambiguity in the formation origin comes from the high survivability of high MMW atmospheres.
% While H$_2$ atmospheres are vulnerable to photoevaporation by young MS stars and hot WDs, high MMW atmospheres on first-generation rocky planets can survive more easily. If first-generation atmospheres survive post-MS evolution, they are indistinguishable from atmospheres on second-generation planets. 
Both Jeans escape and energy-limited hydrodynamic escape have sensitive dependence on molecular weight.
% \footnote{Intense stellar flares of M dwarfs may desiccate or erode high MMW atmospheres via another escape mechanism -- ion escape \citep[e.g.,][]{Airapetian_2017ApJ...836L...3A, Dong_2017ApJ...847L...4D, egan_stellar_2019}. However, WDs cannot have originated from M dwarfs because the MS lifetimes of M dwarfs are longer than the age of the universe.}
The Jeans escape flux $\Phi_J \propto e^{-\lambda_J}$, where $\lambda_J$ is the Jeans escape parameter that is linearly dependent on molecular mass \citep[e.g.,][]{catling_kasting_2017}. Therefore, an 14-fold increase of MMW from mass of H$_2$ (2 amu) to N$_2$ (28 amu) would result in a $e^{-14}$ ($\sim 10^{-7}$) decrease in Jeans escape flux. The decrease would be more drastic for CO$_2$ (44 amu). Energy-limited escape of hydrogen can drag heavy molecules along with it. H$_2$ mixing ratios in our N$_2$- and CO$_2$-dominated atmosphere models do not exceed $\sim 10^{-3}$, so even in a physically unlikely scenario that all hydrogen atoms are lost, and each H atom drags along a N or C atom, the impact of escape on the whole atmosphere is limited. The exact escape rate of heavy molecules, however, would depend on the mass of the heavy molecule, upward H escape flux, thermospheric temperature, and hydrogen diffusion through a high MMW atmosphere \citep{luger_extreme_2015}. Quantitative constraints on hydrodynamic escape rates of heavy species is therefore beyond the scope of this work.

Even if a high MMW atmosphere is lost during the post-MS evolution, a rocky WD planet can potentially replenish the atmosphere via volcanic outgassing, adding another layer of uncertainty to the planet’s origin. 
% Unlike volcanic emission of H$_2$, which is only the dominant form of hydrogen outgassing when the mantle is young and highly reduced, 
Emission of oxidizing heavy molecules can continue at high rates for geological timescale. CO$_2$, for example, can be recycled from the atmosphere to sustain long-term high CO$_2$ emission rates on an active lid planet \citep[e.g.,][]{kasting_evolution_2003}. 
% We assume a $3 \times 10^{11}$ cm$^{-2}$ s$^{-1}$ CO$_2$ emission rate for the high outgassing models, following \cite{hu_photochemistry_2012}. 
N$_2$ is also produced from volcano arcs and mid-ocean ridges on Earth at a rate of $\sim 10^8$ cm$^{-2}$ s$^{-1}$ \citep{fischer_fluxes_2008}, and the geological nitrogen cycle appears to be well balanced by the interplay between volcanic emission and sedimentary burial \citep{catling_kasting_2017}.
% The geological nitrogen cycle appears to be well balanced by the interplay between volcanic emission and sedimentary burial. Biological and anthropogenic activities also recycle atmospheric N$_2$, but their effects should be minor given the relatively small fraction of nitrogen stored in living biomass \citep{catling_kasting_2017}. 
% Because both CO$_2$ and N$_2$ can be recycled to sustain continuous emissions over geological timescale, 
We therefore conclude that high MMW atmospheres dominated by N$_2$ and CO$_2$ can potentially be replenished after erosion.

Even though N$_2$- and CO$_2$-dominated atmospheres cannot uniquely constrain the evolutionary history of a rocky WD exoplanet, these high MMW atmospheres can be conclusively detected by JWST with a small number of transits. For most scenarios, even 1 transit with NIRSpec Prism can rule out a flat spectrum and hence imply the presence of an atmosphere with high significance (Table \ref{table:sigma_flat_line}). Assuming high outgassing rates, flat spectra can be ruled out conclusively for N$_2$ 5000 K WD, N$_2$ 6000 K WD, CO$_2$ 4000 K WD, and CO$_2$ 5000 K WD scenarios at $\gtrsim 5\,\sigma$. For the low outgassing models, a barren rock planet can be ruled out at $> 3\,\sigma$ for all cases, despite weaker spectral features due to reduced emissions. We note two exceptions -- the high outgassing N$_2$ 4000 K WD and CO$_2$ 6000 K WD models have low significance of 2.7 and $2.3\,\sigma$, respectively. This deviation from the general trend can be explained by stochastic noises added to each PandExo simulation. Our simulated JWST observations represent a set of \textit{random} samples, rather than a set of \textit{typical} samples. Fluctuations in calculated significance are therefore normal.

With 5 NIRSpec Prism transits, conclusive ($> 5\,\sigma$) detection is achievable for almost all scenarios for a hypothetical rocky planet orbiting WD 1856+534 in the WD HZ. There are two exceptions, N$_2$ 5000 K WD and CO$_2$ 4000 K WD scenarios, where the significance of ruling out a flat spectrum with 5 transits is lower than the significance with 1 transit. Because other models with similar transit depths have very high significance of ruling out a flat spectrum ranging from 6.5 to $7.8\,\sigma$, we conclude that these two cases represent the pessimistic end of stochastic distribution of PandExo noises. For a typical observation, JWST would be able to conclusively detect a N$_2$ or CO$_2$ atmosphere with 5 or less transits.

%%======================================================================================
%%  DISCUSSION
%%======================================================================================

\section{Discussion} \label{sec:discussion}

\subsection{H$_2$ Escape Mechanisms} \label{sec:511_h2_escape}
% The ``evaporation valley'' in the radius distribution of Kepler exoplanets imply that roughly Earth-mass rocky planets around MS stars cannot retain hydrogen envelopes with $\sim 1$ bar surface pressures. There are some exceptions. It is hypothesized that if the atmospheric mass loss is core-powered, a massive super-Earth may retain some hydrogen envelope \citep[e.g.,][]{gupta_sculpting_2019}. 
% % Indeed, LHS 1140 b, a nearby super-Earth orbiting a M dwarf, lies in the ``evaporation valley'' with a radius of $\sim 1.7\,R_\oplus$ \citep{Ment_2019AJ....157...32M}. 
% Indeed, \cite{owen_hydrogen_2020} argued that we may have already discovered three $\sim2\,M_\oplus$ Kepler planets that fall into this this so-called ``super-puff'' category. The existence of $\sim1\,M_\oplus$ planets with substantial H$_2$ atmospheres, however, lacks observational evidence.
% % planets with large hydrogen- and helium-rich envelopes. 
% % Three Kepler planets with $\sim2\,M_\oplus$ and $\gtrsim 2.5\,R_\oplus$ (Kepler-11f, Kepler-51b, and Kepler-177b) fall into this so-called ``super-puff'' category. 

Atmospheric loss driven by red giant stellar winds has been quantitatively studied for $\sim 0.5\,M_\oplus$ exomoons, Earth-mass planets, and $> 5\,M_\oplus$ super-Earths \citep{ramirez_habitable_2016}. The authors considered Earth-like high MMW atmospheres and concluded that $\sim 100\%$ atmospheric loss will occur for planets receiving Mars-like stellar radiation orbiting F5 or later type stars, when the host stars evolve into red giants. For K5 or later type MS progenitors, even a Saturn equivalent separation cannot prevent total atmospheric loss. Due to the complex nature of turbulent mixing, efficiency of atmospheric loss due to interactions with stellar winds is not well constrained, so we leave quantitative study of H$_2$ atmosphere survivability around red giants for future investigation.

Note that present-day WDs cannot have originated from M dwarfs, because MS lifespans of M dwarfs are considerably longer than the age of the universe. Therefore, escape mechanisms specific to M dwarfs such as ion escape \citep[e.g.,][]{Airapetian_2017ApJ...836L...3A, Dong_2017ApJ...847L...4D} do not apply to WD exoplanets.

\subsection{H$_2$ Production and Retention Mechanisms} \label{sec:512_h2_production}
Because rates of volcanic H$_2$ emissions have been quantitatively discussed in Section \ref{sec:31_h2_indicator}, here we focus on outgassing during accretion \citep[e.g.,][]{elkinstanton_ranges_2008}. A modeling study on various types of common meteorites have shown that a substantial H$_2$-dominated atmosphere can be outgassed via the reaction between water and metallic iron, if sufficient water is added to reduced meteoritic materials \citep{elkinstanton_ranges_2008}. Further studies on mixtures of meteoritic materials showed that H$_2$ would be the dominant form of hydrogen outgassing when oxygen fugacity (a measurement for rock oxidation state) of the mixture is roughly lower than or equal to the oxidation state of O chondrites \citep[e.g.,][]{schaefer_chemistry_2010}.

Compositional study of WD pollutants, which represent the building blocks for second-generation WD exoplanets, implies that second-generation rocky WD planets are capable of outgassing substantial H$_2$. WD pollutants are similar to solar system meteorites in bulk elemental composition and oxygen fugacity \citep{doyle_oxygen_2019}. Measured oxygen fugacity of planetary materials accreted by six polluted WDs are consistent with C chondrites, O chondrites, bulk Earth, and bulk Mars. This compositional similarity implies that at least some second-generation planets formed from WD debris disks have Earth- and Mars-like geophysical and geochemical properties \citep{doyle_oxygen_2019}. Both Earth and Mars are thought to have had an early H$_2$-rich phase \citep[e.g.,][]{ramirez_warming_2014}, so it is not surprising if a geochemically similar second-generation rocky WD planet is born with a hydrogen envelope.

Addition of water to a forming second-generation rocky WD planet will facilitate H$_2$ emission \citep{elkinstanton_ranges_2008}, and water is indeed available in large quantities in WD debris disks. Observations of polluted WDs have shown that accreted materials can consist of 20\% or more water by mass \citep[e.g.,][]{Farihi_2016MNRAS.463.3186F}, in agreement with model predictions that water in minor planets or exomoons can survive post-MS evolution in large quantities \citep[e.g.,][]{malamud_post-main-sequence_2017-1}. Accretion of such water-rich minor planets or comets onto a second-generation WD planet may yield a water reservoir equivalent to $10^{-5}$ to $10^{-2}$ times the mass of Earth’s ocean \citep{Veras_2014MNRAS.445.4175V, vanlieshout_exoplanet_2018}. The initial water content of a second-generation WD planet may be enhanced if it accreted a tidally disrupted water-rich super-Earth. How much water can a second-generation planet recycle from the water-rich debris disk, however, remains an open question.

\cite{kite_superabundance_2019, kite_atmosphere_2020} proposed that the abundance of $2$--$3\,R_\oplus$ sub-Neptunes can be explained by H$_2$ dissolving into long-lived magma ocean under excessive pressure. When H$_2$ atmospheres suffer from atmospheric loss on such planets, hydrogen exsolution from the magma ocean can increase the survivability of H$_2$ envelopes. Therefore, our results cannot be directly adopted to indicate the second-generation origin of H$_2$-rich sub-Neptunes. Nevertheless, planets considered by \cite{kite_superabundance_2019, kite_atmosphere_2020} are in a completely different physical regime ($\gtrsim 4\,M_\oplus$ planets with $\gtrsim 400$ K equilibrium temperature) compared to the temperate $1\,M_\oplus$ rocky planets we consider, so magma-atmosphere interaction does not affect our results.

\subsection{Origin of Close-in Orbits of WD Exoplanets} \label{sec:52_origin_closein}
We consider planets that receive similar irradiation as the Earth. Due to the low luminosity of WDs, these planets need to be in or near the WD HZ ($\sim 0.005$--$0.01$ au) to receive sufficient heating \citep{agol_transit_2011}. The origin of such close-in orbits needs to be accounted for.

Second-generation WD planets naturally have close-in orbits because they form at the 2:1 mean motion resonance with the WD Roche limit. Coincidentally, the 2:1 resonance fits in the HZ of $T_{\rm eff} \lesssim 6000$ K WDs \citep{vanlieshout_exoplanet_2018}.

First-generation planets, however, must be delivered into a close-in orbit by some migration mechanism, because any first-generation planets born close to the star would be engulfed in the red giant phase. There are two mechanisms that can deliver a remote planet to close-in orbits around WDs, namely binary star system interactions and multiplanet scattering.
% -- binary star system interactions \citep[e.g.,][]{Bonsor_2015MNRAS.454...53B, Hamers_2016MNRAS.462L..84H, Stephan_2017ApJ...844L..16S} and multiplanet scattering \citep[e.g.,][]{veras_detectable_2015, mustill_unstable_2018}. 
In a wide binary stellar system, the distant stellar companion can perturb the orbit of a major planet orbiting the evolving post-MS star and potentially deliver it into the Roche limit of its host. Under this mechanism, Neptune-like planets or Kuiper Belt analog objects would spend the first $\sim 200$ Myr of the WD lifetime of its host on a $\sim 100$ au orbit. Subsequently, the planet can be delivered to a close-in orbit with $\sim 10^{-2}$ au periapsis by relatively rapid inward migration that takes $\sim 5$ Myr \citep{Stephan_2017ApJ...844L..16S}. Alternatively, in a closely packed multiplanet system, due to the chaotic nature of scattering, a planet can remain on a $\sim 10$ au orbit for $> 10$ Gyr and then be scattered onto a highly eccentric orbit with periapsis distance only a few percent of an au \citep{veras_detectable_2015}. Tidal interactions with the WD can then circularize its orbit in under $10^3$ years. Efficient tidal interaction also means that planets in WD HZ are tidally locked \citep[e.g.,][]{agol_transit_2011}.

% First-generation planets born close to the star, such as Mercury and Venus, cannot survive the red giant phase because the star will expand to $\sim 1$ au, engulfing the inner planets \citep[e.g.,][]{Schroder_2008MNRAS.386..155S}. Therefore, a first-generation rocky exoplanet orbiting a WD on a tight orbit must have formed further from the star and migrated inwards during the post-MS system evolution. The planet may start out as a major planet, or even as an exomoon of an outer giant planet \citep{ramirez_habitable_2016, veras_post-main-sequence_2016, payne_fate_2017, oconnor_high-eccentricity_2020}.

\subsection{Clouds and Hazes in WD Exoplanet Atmosphere} \label{sec:53_haze}
Pervasive high cloud decks pose a major challenge to the detection of spectral features, because molecules below cloud decks are not accessible. Recent general circulation model (GCM) simulations concluded that transit observations of tidally locked M dwarf terrestrial exoplanets, especially water-rich planets, would be strongly affected by clouds \citep{Komacek_2020ApJ...888L..20K, Suissa_2020ApJ...891...58S}. In the moist greenhouse regime, however, highly saturated stratosphere improves the detectability of H$_2$O despite thick tropospheric clouds \citep{chen_habitability_2019}. Clouds also affect our photochemistry models indirectly by controlling the atmospheric energy budget and by influencing the amount of UV reaching the lower atmosphere. Planets in WD HZ are rapid rotators with $\approx4$--32 hr period \citep{agol_transit_2011}, in contrast to the slowly rotating ($\gtrsim12$ days) M dwarf planets on which cloud effects are most pronounced \citep{Komacek_2020ApJ...888L..20K}.

Our photochemistry models imply that strong haze formation may be prevalent in the atmospheres of WD exoplanets. In an Archean Earth-like N$_2$-dominated anoxic atmosphere, a CH$_4$/CO$_2$ ratio of above 0.1 would lead to strong hydrocarbon haze production driven by CH$_4$ photolysis \citep[e.g.,][]{arney_pale_2016}. Indeed, our photochemistry results show that in N$_2$-dominated models with high outgassing rates, CH$_4$/CO$_2$ equals 0.72, 1.0, and 5.9 for 6000 K, 5000 K, and 4000 K WDs, respectively. In N$_2$-dominated models with low outgassing rates, CH$_4$/CO$_2$ ratios are $6.2 \times 10^{-6}$, $1.4 \times 10^{-6}$, and 1.0 for 6000 K, 5000 K, and 4000 K WDs, respectively. Given the high CH$_4$/CO$_2$ ratios for all high outgassing scenarios and the 4000 K low outgassing scenario, extensive organic haze production in the atmospheres of WD exoplanets may be common.

Even though the production of organic hazes in CO$_2$- and H$_2$-dominated atmospheres is not well studied, CH$_4$/CO$_2$ ratios in our photochemistry models for these types of atmospheres often exceed 0.1. In a CO$_2$-dominated atmosphere with high outgassing rates, CH$_4$/CO$_2$ equals $4.3 \times 10^{-8}$, $8.1 \times 10^{-6}$, and 0.13 for 6000 K, 5000 K, and 4000 K WDs, respectively. In a H$_2$-dominated atmosphere with high outgassing rates, CH$_4$/CO$_2$ equals 0.13, 41, and $2.3 \times 10^4$ for 6000 K, 5000 K, and 4000 K WDs, respectively. If haze production is also positively correlated with CH$_4$/CO$_2$ ratio in highly reducing or highly oxidizing atmospheres, the high CH$_4$/CO$_2$ ratios for H$_2$-dominated models and for the 4000 K WD CO$_2$-dominated model imply that organic hazes on WD exoplanets are common regardless of atmospheric oxidation state. 
% Note that although higher CH$_4$/CO$_2$ ratio means higher likelihood of organic haze production, photochemistry involving sulfur gases can produce organic haze even if the CH$_4$/CO$_2$ ratio is low, given that the planet has high biogenic fluxes of CS$_2$, OCS, CH$_3$SH, and CH$_3$SCH$_3$ \citep{arney_organic_2018}.

Scattering from hazes can complicate detectability of spectral features, but the presence of hazes also provides a new approach to characterize the atmospheres of WD exoplanets and can even indicate biological activity. At visible wavelengths, scattering from hazes produce a slope at $\sim 0.5\,\micron$ and shorter wavelengths \citep[e.g.,][]{Marley_2013cctp.book..367M}, which has a similar shape compared to the Rayleigh scattering slope. In an atmosphere with Earth-like concentration of oxygenic species, such haze slopes may also obscure the O$_3$ Chappuis band at 0.6 $\micron$, complicating the detection of the CH$_4$ + O$_3$ biosignature pair \citep[e.g.,][]{lin_differentiating_2021}. In infrared wavelengths, organic haze has a feature at approximately 6 $\micron$, just outside of the bandpass of NIRSpec Prism \citep{arney_pale_2016, arney_pale_2017}. If organic haze features are detected on a WD exoplanet with low CH$_4$/CO$_2$ ratio, it is possible that the planet has large emissions of biogenic sulfur gases \citep{arney_organic_2018}.

% Diverse production and removal rates of hazes under different atmospheric conditions add another layer of complexity of modeling hazes in an exoplanet atmosphere. Production rates of photochemical hazes have been measured in the laboratory for a wide array of atmospheric conditions, including CO$_2$- and H$_2$-dominated atmospheres \citep{horst_haze_2018, he_photochemical_2018, he_haze_2020}. Removal rates of hazes depend on surface energy of haze molecules, as well as on atmospheric dynamics and precipitation \citep{Yu_2021NatAs...5..822Y}. Haze properties of a WD exoplanet therefore require studies on a case-by-case basis.

\subsection{Habitability of WD Exoplanets} \label{sec:54_habitability}
% Even though our models assume abiotic emissions, Earth-mass WD exoplanets are potentially habitable and are among the most favorable targets for atmospheric characterization including biosignature detection. 
Here we discuss how the changing luminosity, UV radiation, volatile reservoir, and orbital dynamics would affect the habitability of WD exoplanets.

The slow cooling process of WDs provides stable environments for planets around them for billions of years. \cite{agol_transit_2011} estimated that a planet can stay in the continuously HZ of WDs for $> 3$ Gyr. \cite{kozakis_uv_2018} showed that the maximum time a planet spends within the HZ of a 0.6 M$_\odot$ WD is $\sim 6$ Gyr, assuming a conservative HZ defined by the runaway greenhouse and maximum greenhouse effects of H$_2$O and CO$_2$ \citep[e.g.,][]{Kasting_1993Icar..101..108K}. 
% If assuming the empirical HZ where the inner and outer boundaries are defined by irradiation received by recent Venus and early Mars \citep{Kasting_1993Icar..101..108K}, this maximum continuously habitable time can be extended to $\sim 8.5$ Gyr. 
Furthermore, cool single WDs are photometrically stable \citep{Fontaine_2008PASP..120.1043F}, which rules out high energy flares that may cause atmospheric erosion. Flares, which impact habitability of M dwarf planets, should not affect WD planets we consider.

WDs remain cool and quiescent during most of their lifetime, providing a stable low UV environment for any planet orbiting it. High energy UV photons can potentially erode atmospheres and place the water reservoir at risk, due to the photodissociation of water and subsequent hydrogen escape \citep[e.g.,][]{Airapetian_2017ApJ...836L...3A, Dong_2017ApJ...847L...4D}. Some studies have suggested that high UV flux may be necessary for triggering complicated prebiotic chemistry reactions, which are essential for the emergence of life \citep[e.g.,][]{fossati_habitability_2012, Ranjan_2018AsBio..18.1023R}. WDs with $T_{\rm eff} = 6000$ K have UV radiation comparable to the modern Sun (Figure \ref{fig:stellar_spectra}), implying that younger, hotter WDs can output UV photons on the same levels as the young Sun, which is presumed to trigger abiogenesis on Earth.

A rocky WD exoplanet can have an abundant volatile reservoir, which is another requirement for habitability. First-generation rocky WD planets may require either some migration mechanism that delivers the planet from an outer orbit into WD HZ \citep[e.g.,][]{veras_detectable_2015}, or a large initial water fraction. In the latter case, steam-dominated atmospheres on water-rich ``super-Venuses'' can survive for geologic timescale even when exposed to intense EUV fluxes \citep{harman_snowball_2021}. Analogous to planets with high MMW atmospheres considered here, transiting WD planets with steam atmospheres can be readily characterized by JWST, offering an opportunity to study the volatile evolution on terrestrial planets. For a second-generation WD planet, whether it has sufficient volatiles depends on whether the WD debris disk from which the planet formed is volatile-rich. As discussed in Section \ref{sec:512_h2_production}, volatiles such as H$_2$O are indeed abundant in WD debris disks, according to WD pollution observations.

Short tidal circularization and tidal locking timescales imply that planets in the WD HZ are expected to be tidally locked, with orbital periods of $\approx4$--32 hr \citep{agol_transit_2011}. Rapid rotation allows for efficient heat redistribution, preventing nightside atmospheric collapse and hence increasing habitability. Rapid rotation also leads to the formation of narrow global cloud bands, in contrast to thick substellar cloud decks on slowly rotating planets, which have secondary effects on habitability that are explored by GCMs \citep[e.g.,][]{Yang_2014ApJ...787L...2Y}.

\subsection{Future Opportunities} \label{sec:55_future_opportunity}
To date, no Earth-mass transiting WD planet has been discovered. Earth-sized rocky exoplanets transiting a cool WD in the HZ have transit durations of only $\sim 2$ minutes, so detection of such planets require high-cadence observations. Some have searched for such planets using both ground- and space-based facilities \citep[e.g.,][]{van_sluijs_occurrence_2018}. In addition, the feasibility of discovering WD exoplanets has been studied for astrometric detection by Gaia \citep{perryman_astrometric_2014} and for large-scale survey detection by the Large Synoptic Survey Telescope (LSST, \citealt{cortes_detectability_2019}). 
% Microlensing is capable of detecting Jupiter-like planets around WDs \citep{Blackman_2021Natur.598..272B}. Gravitational wave detection of WD exoplanets with the Laser Interferometer Space Antenna (LISA) is also possible, albeit this detection method favors the most massive ($\gtrsim 50\,M_\oplus$) exoplanets \citep{danielski_circumbinary_2019, tamanini_gravitational-wave_2019}. 
The new 20 s cadence mode available during TESS extended mission will provide another avenue of detecting transiting WD exoplanets with short transit durations.

The search for Earth-mass transiting WD exoplanets may be a fruitful one. 
% Assuming a similar occurrence rate of terrestrial exoplanets in WD HZ as in M dwarfs HZ, and assuming a transit probability of $\sim 1\%$, 
About 5 Earth-mass transiting WD exoplanets are expected to be detected within the characterization horizon of JWST, assuming Earth-like atmospheres (\linktocite{kaltenegger_white_2020}{Kaltenegger, MacDonald et al.\ 2020}). For H$_2$ atmospheres, the characterization horizon may be extended. The half-sky survey by LSST may produce a more optimistic number of $\sim 100$ transiting WD exoplanets \citep{cortes_detectability_2019}, although some of those targets may be too dim for JWST to characterize.

%%======================================================================================
%%  CONCLUSION
%%======================================================================================

\section{Conclusion} \label{sec:conclusion}
In this work, we present a photochemical modeling exploration of Earth-mass exoplanets transiting WDs under the context of WD system evolution. We present 1D photochemistry models coupled with an analytical climate model, simulated transmission spectra, and JWST observation models for three types of anoxic atmospheres with different oxidation states. We show that detection of a H$_2$-dominated thick ($\sim 1$ bar) atmosphere indicates a second-generation WD rocky planet, while the detection of a high MMW (N$_2$- or CO$_2$-dominated) atmosphere is degenerate. Detecting H$_2$O, H$_2$, CO$_2$, CO, and CH$_4$ features in H$_2$ atmospheres with JWST requires only 1 transit with NIRSpec Prism. For N$_2$- and CO$_2$-dominated atmospheres, $\sim 25$ NIRSpec Prism transits are required to conclusively detect the above molecules. Buildup of CH$_4$ via photochemical runaway is observed in most models, an effect that can be used to differentiate between high and low outgassing scenarios with 1 JWST transit for H$_2$ atmosphere models and with 25 JWST transits for most N$_2$ and CO$_2$ atmosphere models. For more details on molecule detectability in H$_2$-dominated atmospheres, see Appendix \ref{sec:ap_h2_features}. For additional results and figures for N$_2$- and CO$_2$-dominated atmospheres, see Appendix \ref{sec:add_results_n2co2}.

Earth-mass transiting WD exoplanets are among the most favorable targets for atmospheric characterization of terrestrial planets via transmission spectroscopy. Besides an opportunity to search for biosignature gases on inhabited WD exoplanets, here we show that there is also a ``white dwarf opportunity'' for constraining the evolutionary history of abiotic and prebiotic rocky WD exoplanets. Rocky exoplanets transiting WDs are yet to be found. We intend for the results here to motivate the search for these unique worlds circling dead stars and follow-up atmospheric reconnaissance by JWST.

\begin{acknowledgments}
We thank the anonymous reviewer for constructive comments that improved the clarity of this paper. We thank Sujan Sengupta for co-developing a hydrogen escape code that contributed to this study. Z.L. acknowledges support from the MIT Presidential Fellowship. S. R. thanks Northwestern University for support via the CIERA Postdoctoral Fellowship.
\end{acknowledgments}



%%======================================================================================
%%======================================================================================
%%======================================================================================
%%  APPENDIX
%%======================================================================================
%%======================================================================================
%%======================================================================================

\appendix

%%======================================================================================
%% A - Model Details
%%======================================================================================

\section{Model Details} \label{sec:ap_model_details}

\subsection{Photochemistry Model} \label{sec:ap_chem_details}

Our photochemistry model includes $>800$ chemical reactions, photochemical processes, and emission and thermal escape mechanisms. The model also solves chemical-transport equations for 111 O, H, C, N, and S species, as well as S$_8$ and H$_2$SO$_4$ aerosols, linked by 645 bimolecular reactions, 85 ter-molecular reactions, and 93 thermal dissociation reactions (see \citealt{hu_photochemistry_2012} for a full list of species and reactions). The reaction network has recently been updated to include nitrogenous chemistry, and the rate laws of several reactions have been updated as well (see \citealt{ranjan_photochemical_2022} for details). Note that because photochemistry of higher hydrocarbons and organic haze production involve many uncertainties, \cite{hu_photochemistry_2012} excluded reactions involving molecules containing more than two carbon atoms in an ad hoc fashion by assuming a high ($10^{-5}$ cm s$^{-1}$) deposition velocity for C$_2$H$_6$. Implications of hazy atmospheres are discussed in Section \ref{sec:53_haze}.

For each atmospheric oxidation state, we use a subset of the full reaction network that is relevant. We assume the planets are covered with a substantial surface liquid water ocean and water vapor is transported upwards at a constant flux of $10^{-2}$ cm$^{-2}$ s$^{-1}$ due to evaporation. We consider a zero rainout rate to simulate an ocean that is saturated with H$_2$, CO, CH$_4$, C$_2$H$_6$, and O$_2$ on an abiotic planet \citep[following][]{hu_photochemistry_2012}. Dry deposition velocities assumed in our model follow the exoplanet benchmark cases parameters (Table 5, \citealt{hu_photochemistry_2012}). The photochemical model is considered to be converged when the variation timescale of each species at each altitude exceeds the diffusion timescale of the entire atmosphere. Key model parameters are summarized in Table \ref{table:model_parameters}.

Our high and low outgassing rates are differ by a factor of 1000. This factor is not arbitrary. Volcanic production rate of modern Earth is $\sim$ 100--1000 times higher than Venus \citep{gaillard_theoretical_2014, gillmann_atmospheremantle_2014}. To explore a wider parameter space, we choose a factor of 1000 instead of 100. 
% We will justify simulating two sets of outgassing rates in Section \ref{sec:31_h2_indicator} under the context of WD exoplanet evolution.
% respectively, to maintain the same 288 K surface temperature for all 6000 K WD models. This approximation is valid despite \cite{hu_photochemistry_2012} assumed solar spectrum while we assumed a 6000 K WD spectrum, because the two stellar spectra have similar energy distribution (Figure \ref{fig:stellar_spectra}).

The required inputs of the temperature model include the planet’s equilibrium temperature, interior temperature, mean thermal opacity of the atmosphere, and mean optical opacity. Equilibrium temperatures of our models are summarized in Table \ref{table:model_parameters}. For all models, we assume an Earth-like interior temperature of 35.7 K, which is calculated based on an estimated Earth’s total surface heat flow of $47 \pm 2 \times 10^{12}$ W \citep{Davies_2010SolE....1....5D}. Mean thermal and optical opacities were calculated based on mixing ratios of key absorbing species, such as CH$_4$, CO$_2$, H$_2$O, and H$_2$, produced by the photochemistry model, in combination with cross section data from the Exoclimes Simulation Platform \citep{grimm_helios-k_2015} and the MPI-Mainz Spectral Atlas \citep{keller-rudek_mpi-mainz_2013}. Opacities calculated from photochemistry model outputs are inputted to the temperature model as initial conditions, while temperature-pressure profile generated by the temperature model is in turn inputted to the photochemistry model. The two models were run iteratively until surface temperature predicted by both models differ by less than 1 K. We run several extra iterations after convergence to ensure stable solution. Surface temperatures for all three 6000 K WD models are fixed at 288 K, and the thermal and optical opacities of the 6000 K WD models were used as benchmark for the 5000 K and 4000 K WD models. The cooler WD models have higher concentration of gases with high thermal opacities due to lower UV radiation from host stars and hence lower photodissociation rate. This would increase the thermal opacities of cooler WD models relative to the 6000 K WD model, causing temperatures to increase. The final converged surface temperatures for all the models are summarized in Table \ref{table:model_parameters}.

%%% TABLE 1 - MODEL PARAMETERS ================
\begin{table*}[tb]
\caption{Model Parameters for the Three Anoxic Atmospheric Compositions}
\label{table:model_parameters}
\centering
\begin{tabular}{llcccccc}
\hline\hline

\multicolumn{2}{l}{\multirow{2}{*}{Parameters}}     & \multicolumn{2}{c}{Reducing}  & \multicolumn{2}{c}{Weakly oxidizing}    & \multicolumn{2}{c}{Highly oxidizing}                   \\
\multicolumn{2}{c}{}  & \multicolumn{2}{c}{(90\% H$_2$, 10\% N$_2$)}      & \multicolumn{2}{c}{($> 99$\% N$_2$)}    & \multicolumn{2}{c}{(90\% CO$_2$, 10\% N$_2$)}                    \\
\hline

\multicolumn{2}{l}{Outgassing scenario}   & \multicolumn{1}{c}{high}          & \multicolumn{1}{c}{low} &
\multicolumn{1}{c}{high}          & \multicolumn{1}{c}{low} & \multicolumn{1}{c}{high}          & \multicolumn{1}{c}{low}                           \\
\hline

\multicolumn{2}{l}{Equivalent semi-major axis (au)} & 1.6 & 1.6 & 1.0 & 1.0 & 1.3 & 1.3 \\
\multicolumn{2}{l}{Mass ($M_\oplus$)}   & \multicolumn{1}{c}{1.0} & \multicolumn{1}{c}{1.0} & \multicolumn{1}{c}{1.0} & \multicolumn{1}{c}{1.0} & \multicolumn{1}{c}{1.0} & \multicolumn{1}{c}{1.0}  \\
\multicolumn{2}{l}{Radius ($R_\oplus$)} & \multicolumn{1}{c}{1.0} & \multicolumn{1}{c}{1.0} & \multicolumn{1}{c}{1.0} & \multicolumn{1}{c}{1.0} & \multicolumn{1}{c}{1.0} & \multicolumn{1}{c}{1.0}  \\
\multicolumn{2}{l}{Surface pressure (bar)}          & 1.0 & 1.0 & 1.0 & 1.0 & 1.0 & 1.0 \\
\hline

\multirow{3}{*}{Surface temperature (K)} & 4000 K WD & 347 & 288 & 299 & 284 & 290 & 288 \\
 & 5000 K WD & 290 & 288 & 289 & 284 & 288 & 288 \\
 & 6000 K WD & 288 & 288 & 288 & 284 & 288 & 288 \\
\hline

\multirow{3}{*}{MMW (amu)} & 4000 K WD & 13 & 4.6 & 28 & 28 & 39 & 42 \\
 & 5000 K WD & 4.7 & 4.6 & 28 & 28 & 42 & 42 \\
 & 6000 K WD & 4.6 & 4.6 & 28 & 28 & 42 & 42 \\
\hline

\multirow{5}{*}{Gas emission (cm$^{-2}$ s$^{-1}$)}            & CO$_2$      & $3\times10^{11}$ & $3\times10^{8}$ & $3\times10^{11}$ & $3\times10^{8}$ & \multicolumn{1}{c}{N/A} & \multicolumn{1}{c}{N/A}                           \\
 & H$_2$       & \multicolumn{1}{c}{N/A} & \multicolumn{1}{c}{N/A} & $3\times10^{10}$ & $3\times10^{7}$ & $3\times10^{10}$ & $3\times10^{7}$ \\
 & SO$_2$      & $3\times10^{9}$ & $3\times10^{6}$ & $3\times10^{9}$ & $3\times10^{6}$ & $3\times10^{9}$ & $3\times10^{6}$ \\
 & CH$_4$      & $3\times10^{8}$ & $3\times10^{5}$ & $3\times10^{8}$ & $3\times10^{5}$ & $3\times10^{8}$ & $3\times10^{5}$ \\
 & H$_2$S      & $3\times10^{8}$ & $3\times10^{5}$ & $3\times10^{8}$ & $3\times10^{5}$ & $3\times10^{8}$ & $3\times10^{5}$ \\
\hline
\multirow{2}{*}{Water and rainout}       & f(H$_2$O) surface emission & 0.01     & 0.01     & 0.01     & 0.01     & 0.01     & 0.01     \\
 & rainout rate            & 0   & 0   & 0   & 0   & 0   & 0   \\
\hline

\end{tabular}
\end{table*}
%%% TABLE 1 - MODEL PARAMETERS ================

\subsection{Transmission Spectra Model} \label{sec:ap_tran_details}
Our transmission spectra model includes the most spectroscopically relevant species: C$_2$, C$_2$H$_2$, C$_2$H$_4$, CH, CH$_3$, CH$_4$, CN, CO, CO$_2$, CS, H$_2$, H$_2$O, H$_2$O$_2$, H$_2$S, HCN, HNO$_3$, NH, NH$_3$, O$_3$, OH, SO$_2$, and SO$_3$. We include the following Rayleigh scattering species: H$_2$O, CO, CH$_4$, CO$_2$, H$_2$, O$_2$, and N$_2$. Collision-induced absorption (CIA) plays an important role in transmission spectroscopy, especially when one or both species in a CIA pair exist in high concentration in the modeled atmosphere. We therefore include the following CIA pairs: N$_2$-N$_2$, H$_2$-H$_2$, O$_2$-O$_2$, N$_2$-O$_2$, CO$_2$-CO$_2$, and H$_2$-He.

Detailed vertical and horizontal distribution of realistic 3D clouds on exoplanets is highly uncertain. For generality, we assume a single homogeneous cloud layer located at 6 km in the N$_2$-dominated atmosphere \citep[following][]{lin_differentiating_2021}. For consistency across atmospheres with different scale heights, we place the cloud layer at 0.47 bar in the CO$_2$- and H$_2$-dominated models, which corresponds to 6 km in the N$_2$-dominated atmospheres. We account for the most pessimistic scenario by assuming the cloud layer is fully opaque and covers the entire day-night terminator.

Effect of atmospheric refraction is dependent on $R_* / a$, the inverse of the scaled semimajor axis, and is strongest when this value is small \citep{betremieux_2014ApJ...791....7B, Robinson_2017ApJ...850..128R}. Applying Equation (14) from \cite{Robinson_2017ApJ...850..128R}, we find that the maximum pressures accessible to transmission spectroscopy are $\approx 0.7$, $\approx 4$, and $\approx 0.3$ bar, for N$_2$, H$_2$, and CO$_2$ atmospheres, respectively, with uncertainties coming from the exact choice of orbital distance, MMW, and atmospheric refractivity. The refractive cutoffs in N$_2$- and H$_2$-dominated atmospheres are below the cloud layer at 0.47 bar and therefore do not affect the transmission spectra. The refractive cutoff in CO$_2$-dominated atmosphere, however, is above the cloud deck, so we introduce a completely opaque layer at 0.3 bar.

%%======================================================================================
%% B - Model Details
%%======================================================================================

\section{Quantitative Analysis of H$_2$-dominated Atmosphere as Indicator of Second-generation Planet} \label{sec:ap_quantitative}

\subsection{Survivability of Hydrogen Atmospheres on First-generation WD Exoplanets} \label{sec:ap_survivability}
In this section, we quantitatively discuss hydrogen escape on a first-generation rocky planet during the hot WD phase. We show that even if the planet migrates to a 50--100 au orbit, excessive EUV radiation from the hot WD can lead to complete erosion of a $\sim1$ bar hydrogen atmosphere.

Two types of neutral atmospheric escapes are relevant for rocky planets, namely Jeans escape and hydrodynamic escape. Hydrodynamic escape dominates at high incident fluxes and can produce extremely high mass loss rates that can account for the evaporation of entire atmospheres \citep{owen_kepler_2013}. We therefore focus on the hydrodynamic escape mechanism and calculate the ``energy-limited'' mass loss rate given the EUV irradiation from a hot WD.

Intense EUV radiation from hot WDs results in extreme mass loss rates. EUV flux from a hot ($\gtrsim 30,000$ K) WD exceeds EUV flux of the young Sun by a factor of $10^2$--$10^3$, while the young Sun is thought to be $10^3$ times more active than modern Sun \citep{Schreiber_2019ApJ...887L...4S}. Even at a distance of $\sim 50$--$100$ au, EUV flux from a 40,000--80,000 K WD can be as strong as $\sim0.1$--$10$ W m$^{-2}$. In this EUV flux range, atmospheric escape is dominated by the so-called ``energy-limited'' escape \citep[e.g.,][]{luger_extreme_2015, sengupta_upper_2016}. Here we adopt equation (2) in \cite{luger_extreme_2015} to calculate the mass loss rate on rocky exoplanets orbiting hot WDs:
\begin{equation}
    \dot{M}_{\rm EL} = \frac{\epsilon_{\rm EUV} \pi \mathcal{F}_{\rm EUV} R_{\rm p} R_{\rm EUV}^2}{G M_{\rm p} K_{\rm tide}}
\end{equation}
where $\mathcal{F}_{\rm EUV}$ is the EUV flux, $R_{\rm p}$ is the planet radius, $R_{\rm EUV}$ is the radius at which the bulk of EUV energy is deposited, $\epsilon_{\rm EUV}$ is the EUV absorption efficiency, and $K_{\rm tide}$ is the tidal correction term. We make several simplifications here: $R_{\rm EUV}$ is assumed to be equal to $R_p$ because scale height $H << R_p$ on a rocky planet, $\epsilon_{\rm EUV}$ is assumed to be 0.3, and $K_{\rm tide}$ is assumed to be 1 \citep[following][]{luger_extreme_2015}. We assume $1\,R_\oplus$ radius and $1\,M_\oplus$ mass. We consider a scenario that the planet migrates to a $\sim 50$--$100$ au orbit around a $\sim 40,000$ K WD, which is optimistic for hydrogen survival. EUV flux received by the planet in this case is assumed to be 0.1 W m$^{-2}$ \citep[following][]{Schreiber_2019ApJ...887L...4S}. Hydrogen escape flux in this scenario is $4.1 \times 10^7$ kg s$^{-1}$, which is equivalent to $2.4 \times 10^{31}$ H atoms per s. In addition, we perform a sensitivity test assuming different values of $\epsilon_{\rm EUV}$ and $R_{\rm EUV}$. \cite{luger_extreme_2015} considered $0.15 \leq \epsilon_{EUV} \leq 0.3$ as typical for H$_2$-rich atmospheres. Our model H$_2$ atmospheres have maximum altitudes of $\sim 1200$ km, and we choose half of the maximum altitude (600 km) as the maximum $R_{\rm EUV}$. Pressures at 600 km in our models are $\sim 10^6$ times lower than Earth’s exobase pressure. The sensitivity test results in a factor of $\approx 2.4$ change in the energy-limited escape rate (Figure \ref{fig:H_emissions_escape}).

A first-generation WD planet can maintain its H$_2$-dominated atmosphere if sources of hydrogen overwhelm sinks. We make a first-order assumption that the sole sink for hydrogen is escape at the top of atmosphere and the sole source is volcanic emission. We consider an optimistic case that H$_2$ outgassing is 20 times higher than modern Earth volcanic emission. The volcanic H$_2$ flux on modern Earth is estimated to be $1.5 \times 10^{9}$ cm$^{-2}$ s$^{-1}$ \citep{james_photochemical_2018}. On young planets with more reduced mantles or planets with additional internal heating due to tidal dissipation, $\sim 20$ times higher production is plausible. We therefore assume $3.0 \times 10^{10}$ cm$^{-2}$ s$^{-1}$ flux as a ``high outgassing'' scenario \citep[following][]{james_photochemical_2018}. We also consider a pessimistic ``low outgassing'' scenario where H$_2$ outgassing is 1000 times lower than the high outgassing scenario. The high and low outgassing rates translates to global H atom production rates of $3.1 \times 10^{29}$ s$^{-1}$ and $3.1 \times 10^{26}$ s$^{-1}$ (Figure \ref{fig:H_emissions_escape}).

An immediate observation from Figure \ref{fig:H_emissions_escape} is that even in the optimistic high outgassing scenario, hydrogen production is significantly less than hot WD EUV-driven hydrogen loss by a factor of $\sim 100$. Therefore, net H$_2$ loss will occur even in a the most optimistic scenario for hydrogen retention, where the planet migrates to a separation of $\sim 50$--$100$ au, orbits a relatively cool ($\sim 40,000$ K) young WD, and emits 20 times more H$_2$ than modern Earth. The mass loss rate in this scenario is $\sim 4\times10^4$ kg s$^{-1}$, or $\sim 10^{12}$ kg yr$^{-1}$. Total mass of Earth’s atmosphere is on the order of $\sim 10^{18}$ kg, so a H$_2$ atmosphere on an Earth-mass first-generation WD exoplanet will be evaporated entirely in a timescale of $\sim 1$ Myr. This timescale is shorter than the $\sim 10$ Myr cooling time for the effective temperature of a 100,000 K WD to drop to 30,000 K \citep{fontaine_potential_2001}. Cooling below 30,000 K is much slower. It takes $\sim 2$ Gyr to cool a WD to 6000 K \citep{Bergeron_2001ApJS..133..413B, fontaine_potential_2001}, and at this temperature the WD still emits EUV radiation comparable to modern Sun (Figure \ref{fig:stellar_spectra}). The extended exposure to high levels of EUV photon flux guarantees total evaporation of hydrogen envelopes on first-generation rocky WD planets.

\subsection{Escape Mechanisms on Rocky WD Exoplanet} 

Here we quantitatively study the escape mechanisms on a rocky planet orbiting cool ($T_{\rm eff} \lesssim 5000$ K) WDs. We conclude that maintaining a H$_2$-dominated atmosphere around a cool WD is possible for planets with high outgassing rates and Earth-like exobase temperature ($T_{\rm exo}$), or planets with low outgassing rates and Venus- or Mars-like $T_{\rm exo}$. We break down our quantitative analysis in the order of Jeans escape, energy-limited escape, and diffusion. H$_2$ emission and escape fluxes are visualized in Figure \ref{fig:H_emissions_escape}.

On a planet with low EUV irradiation, Jeans escape is the dominant escape mechanism. Jeans escape is controlled by the Jeans escape parameter, which is inversely proportional to $T_{\rm exo}$ \citep{Hunten_1973JAtS...30.1481H, tian_hydrodynamic_2008}. At low exobase temperatures ($T_{\rm exo} \lesssim 1000$ K), Jeans escape flux is a very sensitive function of exobase temperature, where a $\sim$ 300 K decrease in $T_{\rm exo}$ can lead to two orders of magnitude decrease in Jeans escape flux.

It should be reasonable to assume that $T_{\rm exo}$ on cool WD exoplanets with anoxic atmospheres are lower than Earth’s $T_{\rm exo}$, despite the lack of model constraints. Earth’s exobase is heated by several mechanisms, including UV-induced photoionization and photodissociation of CO$_2$, N$_2$, O, O$_2$ and O$_3$, where the most efficient heat source among all mechanisms is photoionization and photodissociation of O$_2$ (see e.g., \citealt{kulikov_comparative_2007} and references therein). In anoxic atmospheres, such as CO$_2$-dominated Venusian and Martian atmospheres, $T_{\rm exo}$ can be as low as 275 K and 350 K, respectively \citep{dePater_2001plsc.book.....D}. For comparison, Earth’s exobase is heated to a temperature of $\sim 1000$ K \citep[e.g.,][]{tian_hydrodynamic_2008}. Because exospheric heating is powered by UV, $T_{\rm exo}$ varies with stellar activity. Earth’s exobase temperature ranges from $\approx 700$ K to $\approx 1500$ K depending on solar activity, with some uncertainties from the assumed eddy diffusion coefficient \citep{roble_global_1987}. To explore a wide parameter space, we consider Jeans escape for three $T_{\rm exo}$: 1500 K, 700 K, and 350 K. Note that all three types of atmospheres we consider are anoxic, and the EUV levels of cool WDs are significantly lower than modern Sun, so even the 700 K ``solar min'' exobase temperature is likely to be an overestimate for cool WDs. But we stick with these temperature choices to account for $\gtrsim 6000$ K WDs with roughly solar level UV radiation.

We calculate Jeans escape rates on WD exoplanets based on formalism derived by \cite{Hunten_1973JAtS...30.1481H} and \cite{watson_dynamics_1981}. The Jeans escape rates are shown in Figure \ref{fig:H_emissions_escape} alongside with H$_2$ emission rates for comparison. The high outgassing rate is $\sim 3$ times higher than the ``solar max'' Jeans escape rate and $\sim 100$ times higher than the ``solar min'' Jeans escape rate. The low H$_2$ outgassing rate is lower than the ``solar min'' Jeans escape rate but is $\sim 1000$ times higher than the Jeans escape rate assuming $T_{\rm exo} = 350$ K. The low H$_2$ outgassing rate also is comparable to the energy-limited escape rate around a 5000 K WD, but is many orders of magnitudes higher than the energy-limited escape rate around a 4000 K WD. Given that even the ``solar min'' escape rate is likely an overestimate for cool WDs, we conclude that a rocky planet around cool WDs can maintain a hydrogen atmosphere if it has modern Earth-like or higher H$_2$ emissions. Even a planet with 1000 times reduced H$_2$ emission can maintain a H$_2$-dominated atmosphere around a 4000 K WD. Survival of H$_2$ atmosphere on a low outgassing planet orbiting a 5000 K WD would depend on the EUV levels of the particular WD and heating mechanisms in the planet’s exosphere.

Energy-limited hydrodynamic escape is important when EUV flux is high. In Figure \ref{fig:H_emissions_escape}, we show the energy-limited escape rate as a function of WD EUV radiation, following two analytical models \citep{luger_extreme_2015, sengupta_upper_2016}. Hydrogen escape flux due to energy-limited escape only becomes comparable to Jeans escape for $\gtrsim 6000$ K WDs. Energy-limited escape flux drops logarithmically as EUV flux decreases and is orders of magnitudes lower than Jeans escape flux for $\lesssim 5000$ K WDs and is therefore neglected.

On Earth, hydrogen escape is not limited by the rate at which H atoms diffuse into space, but rather by the rate at which hydrogen is delivered to the exobase from the collisional lower atmosphere \citep[e.g.,][]{catling_kasting_2017}. In an atmosphere where H$_2$ is the dominant constituent, however, supply of hydrogen is practically infinite. Thus, diffusion-limited escape is neglected.

\section{Detectability of Spectral Features in H$_2$-Dominated Atmospheres} \label{sec:ap_h2_features}
Here we discuss the locations and detectability of key spectral features in H$_2$-dominated atmospheres based on Figure \ref{fig:spectra_pandexo_h2} and \ref{fig:individual_species_h2}. The most prominent feature in our model spectra is the H$_2$-H$_2$ CIA extending from $\sim 6$ to 20 $\micron$. Rayleigh scattering slope at $\lesssim 1\,\micron$ is also dominated by H$_2$ opacity, except in the 4000 K high outgassing case, where CH$_4$ becomes pervasive in the atmosphere with a mixing ratio of 0.73. CH$_4$ features at 1.7, 2.3, 3.4, and 7.5 $\micron$ are strong in most scenarios. CO$_2$ features are strong in the high outgassing models, with the 4.2 $\micron$ feature being the most prominent because the 15 $\micron$ feature is partially obscured by the broad H$_2$-H$_2$ CIA wing. H$_2$O features at 2.6 and 6.4 $\micron$ are present in some models, especially in the 6000 K WD models where the CH$_4$ feature at 7.5 $\micron$ is relatively weak and does not completely overlap with the 6.4 $\micron$ water feature. CO shows a strong feature at 4.7 $\micron$, which has comparable strength as the 4.2 $\micron$ CO$_2$ feature and is not obscured by any overlapping stronger features. The potential of detecting CO in large amount (the mixing ratios of CO range from $8.0 \times 10^{-5}$ to $3.7 \times 10^{-4}$ in the high outgassing models) in a H$_2$-dominated atmosphere is noteworthy, because CO is the best-studied species under the context of photochemical runaway \citep[][]{schwieterman_rethinking_2019, ranjan_photochemical_2022}. 
% In addition, some argue that CO is an antibiosignature \citep{ragsdale_life_2004, Wang_2016Icar..266...15W, Catling_2018AsBio..18..709C}, so the 4.7 $\micron$ CO feature may help to constrain the surface habitability of rocky WD exoplanets.

%%======================================================================================
%%  D - ADDITIONAL RESULTS - N2 AND CO2
%%======================================================================================

%%% TRANSMISSION SPECTRA W/ PANDEXO, N2 ================
\begin{figure*}[!p]
\centering
\includegraphics[width=0.7\textwidth]{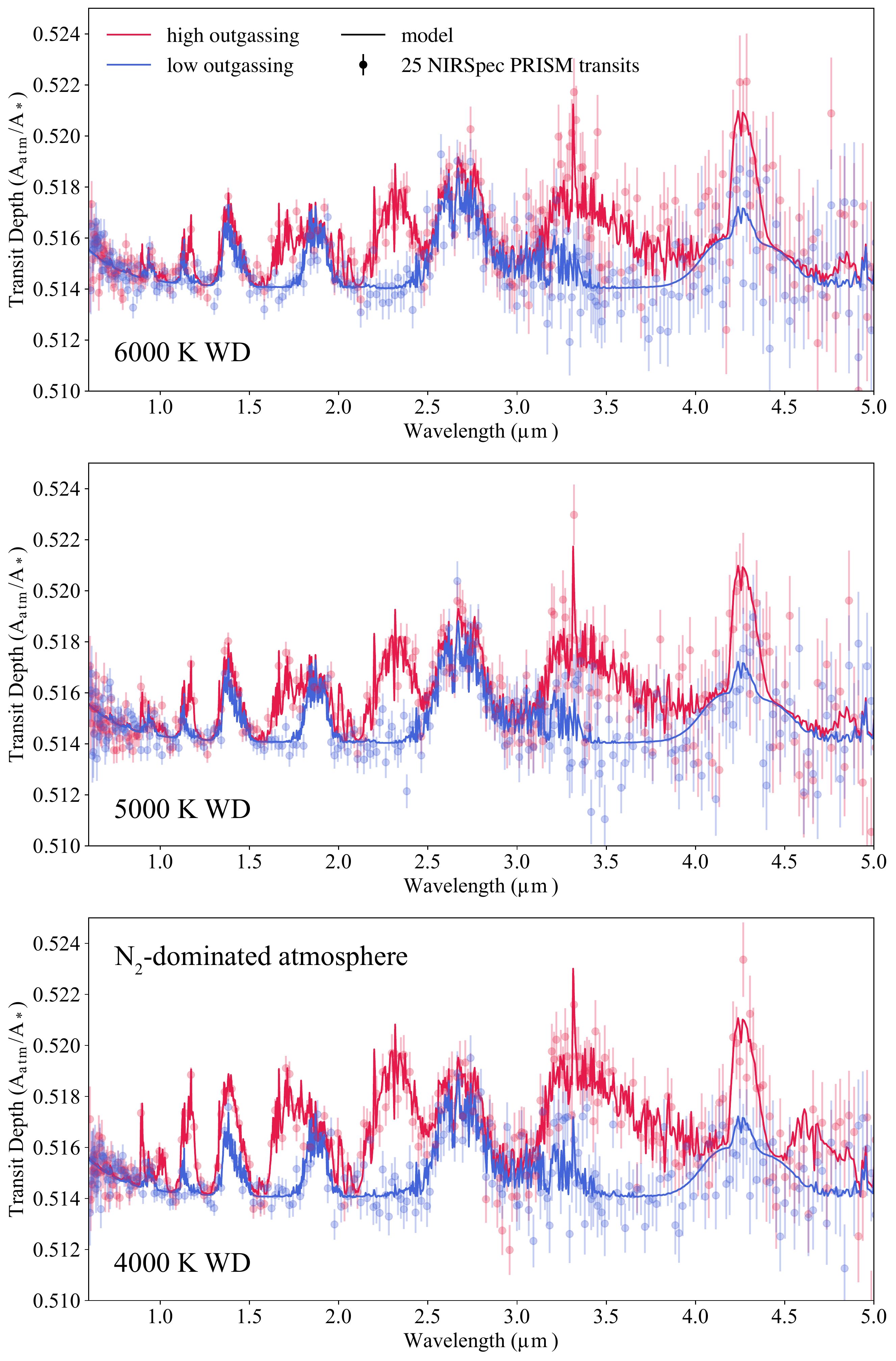}
\caption{Transmission spectra models and simulated JWST transit observations for N$_2$-dominated atmospheres, assuming a $1\,R_\oplus$ planet transiting WD 1856+534. We compare high outgassing rate models (red solid lines) with low outgassing rate models (blue solid lines) and show simulated JWST data points with their $1\,\sigma$ error bars. We assume 25 NIRSpec Prism transits for the N$_2$-dominated atmosphere models. Major spectral features are detectable in the N$_2$-dominated atmosphere models. High and low outgassing scenarios are potentially distinguishable from different strengths of CH$_4$ features in all three models. In the 4000 K WD model, high and low outgassing scenarios are most easily distinguishable due to runaway buildup of CH$_4$.   \label{fig:spectra_pandexo_n2}}
\end{figure*}
% H$_2$-, N$_2$-, and CO$_2$-dominated atmospheres are shown in the top, middle, and bottom rows, respectively. The left, middle, and right columns show WDs with effective temperatures of 6000 K, 5000 K, and 4000 K, respectively. Note that we assume 1 NIRSpec Prism transit for the H$_2$-dominated atmosphere models and 25 NIRSpec Prism transits for the N$_2$-, and CO$_2$-dominated models. In the H$_2$-dominated atmospheres, spectral signatures have high detectability, and differentiating between high and low outgassing scenarios is achievable with 1 transit. With 25 transits, major spectral features are also detectable in the high MMW atmosphere models, but whether the high and low outgassing scenarios are distinguishable depends on whether CH$_4$ accumulates to much higher level in the high outgassing models.
%%% TRANSMISSION SPECTRA W/ PANDEXO, N2 ================

%%% INDIVIDUAL SPECIES N2 ================
\begin{figure*}[t!]
\includegraphics[width=\textwidth]{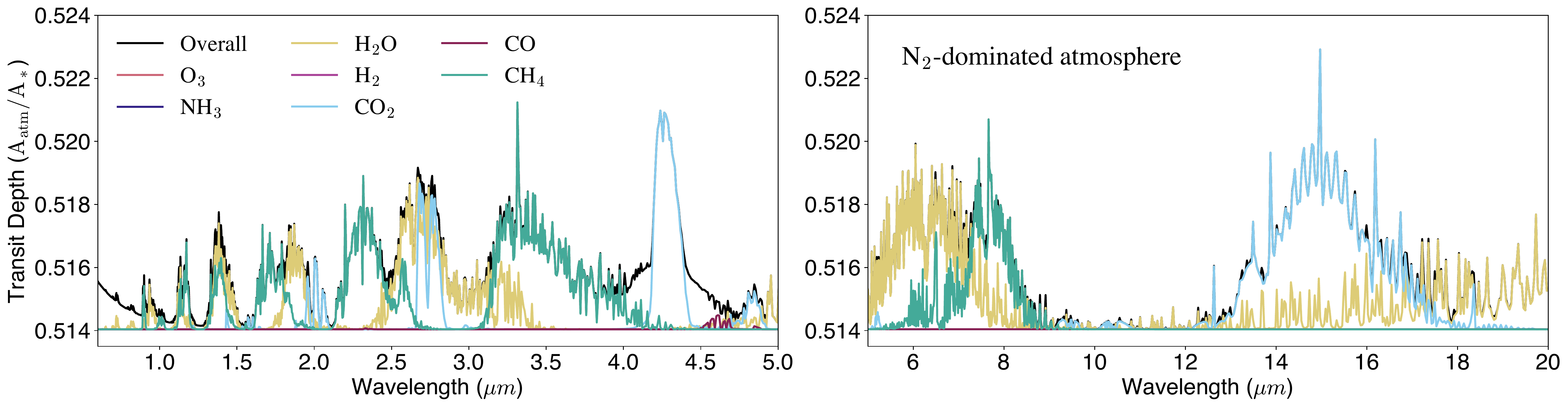}
\caption{Molecular contributions to the overall model transmission spectra for N$_2$-dominated atmospheres. \label{fig:individual_species_n2}}
\end{figure*}
% Molecular contributions to the overall model transmission spectra for H$_2$-, N$_2$-, and CO$_2$-dominated atmospheres (top to bottom). Only planets with high outgassing rates around 6000 K WDs are shown for simplicity. Only features in the bandpass of NIRSpec Prism (0.6-5 $\micron$) are shown to allow comparison with Figure \ref{fig:spectra_pandexo}.
%%% INDIVIDUAL SPECIES N2 ================

%%% TRANSMISSION SPECTRA W/ PANDEXO, CO2 ================
\begin{figure*}[!p]
\centering
\includegraphics[width=0.7\textwidth]{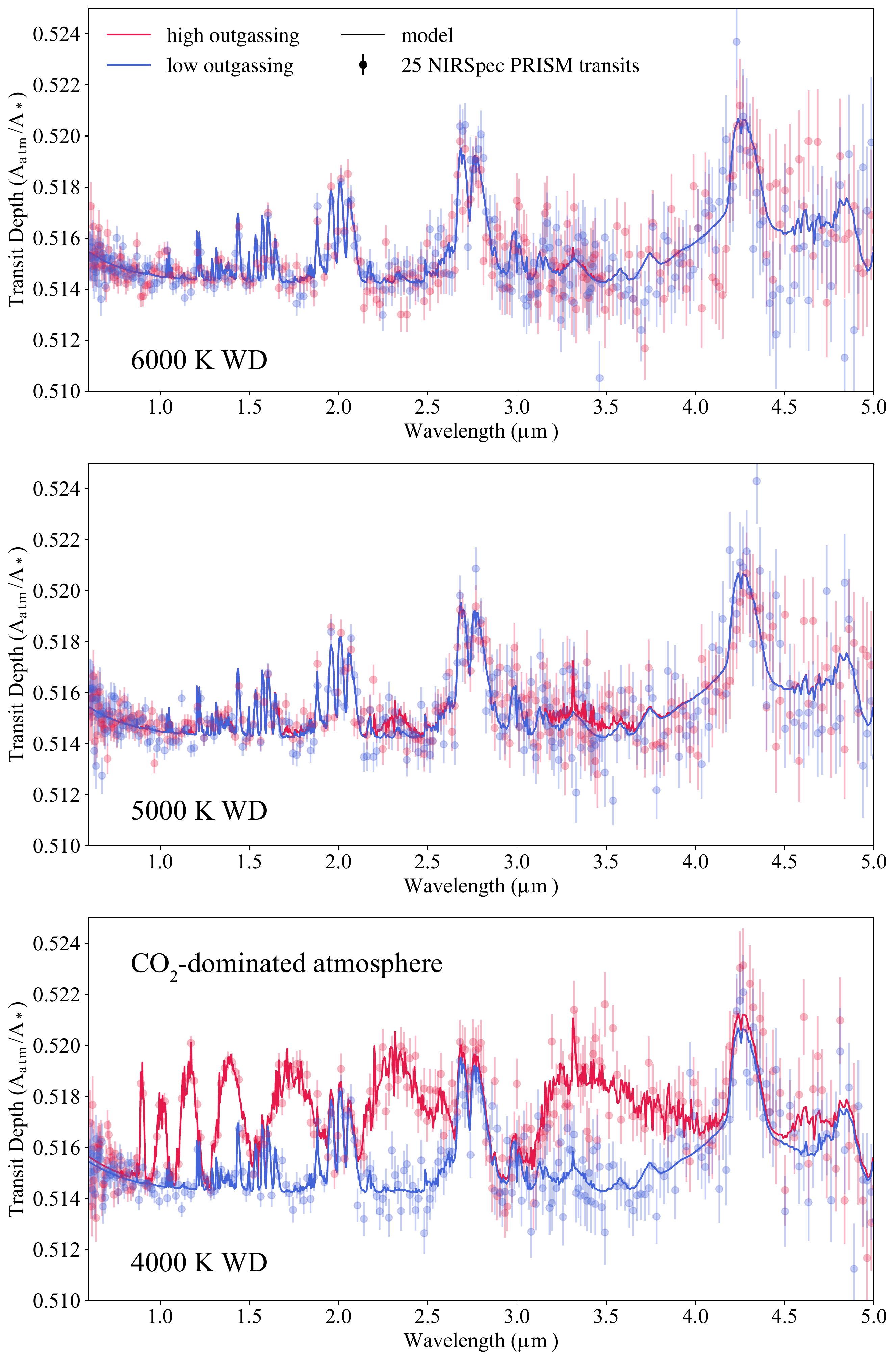}
\caption{Transmission spectra models and simulated JWST transit observations for CO$_2$-dominated atmospheres, assuming a $1\,R_\oplus$ planet transiting WD 1856+534. We compare high outgassing rate models (red solid lines) with low outgassing rate models (blue solid lines) and show simulated JWST data points with their $1\,\sigma$ error bars. We assume 25 NIRSpec Prism transits for the CO$_2$-dominated atmosphere models. CO$_2$ features are detectable with 25 transits for all models, but distinguishing between high and low outgassing scenarios is only achievable for the 4000 K WD model, because CH$_4$ only builds up to detectable level in the 4000 K WD high outgassing scenario.  \label{fig:spectra_pandexo_co2}}
\end{figure*}
% H$_2$-, N$_2$-, and CO$_2$-dominated atmospheres are shown in the top, middle, and bottom rows, respectively. The left, middle, and right columns show WDs with effective temperatures of 6000 K, 5000 K, and 4000 K, respectively. Note that we assume 1 NIRSpec Prism transit for the H$_2$-dominated atmosphere models and 25 NIRSpec Prism transits for the N$_2$-, and CO$_2$-dominated models. In the H$_2$-dominated atmospheres, spectral signatures have high detectability, and differentiating between high and low outgassing scenarios is achievable with 1 transit. With 25 transits, major spectral features are also detectable in the high MMW atmosphere models, but whether the high and low outgassing scenarios are distinguishable depends on whether CH$_4$ accumulates to much higher level in the high outgassing models.
%%% TRANSMISSION SPECTRA W/ PANDEXO, CO2 ================

%%% INDIVIDUAL SPECIES CO2 ================
\begin{figure*}[t!]
\includegraphics[width=\textwidth]{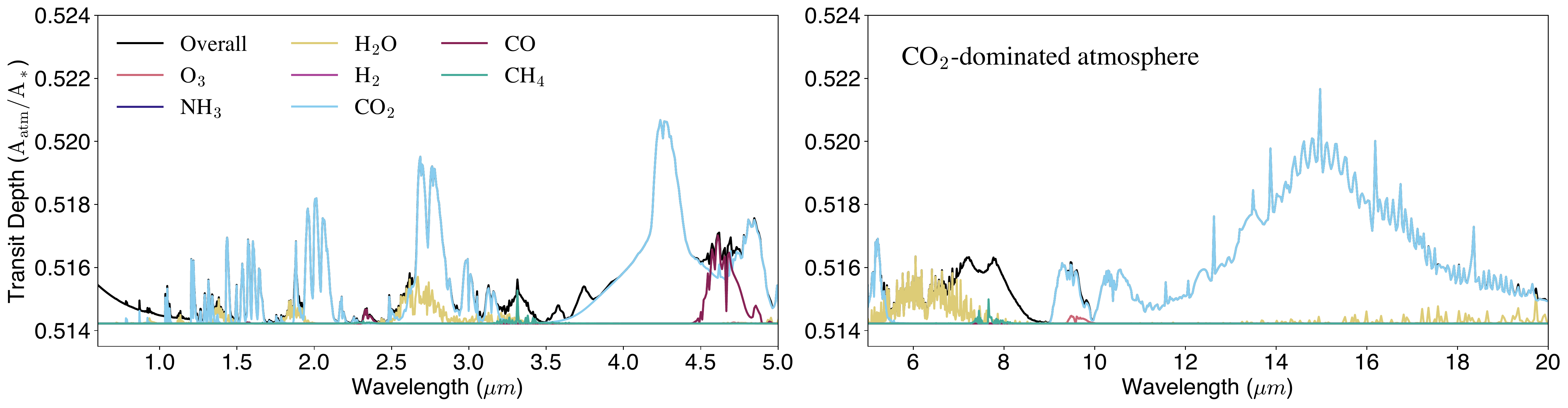}
\caption{Molecular contributions to the overall model transmission spectra for CO$_2$-dominated atmospheres. \label{fig:individual_species_co2}}
\end{figure*}
% Molecular contributions to the overall model transmission spectra for H$_2$-, N$_2$-, and CO$_2$-dominated atmospheres (top to bottom). Only planets with high outgassing rates around 6000 K WDs are shown for simplicity. Only features in the bandpass of NIRSpec Prism (0.6-5 $\micron$) are shown to allow comparison with Figure \ref{fig:spectra_pandexo}.
%%% INDIVIDUAL SPECIES CO2 ================

\section{Additional Results for N$_2$- and CO$_2$-dominated Atmospheres} \label{sec:add_results_n2co2}

\subsection{Inferring Outgassing Activities of Planets with High MMW Atmospheres} \label{sec:42_infer_outgass}

As demonstrated in Section \ref{sec:ap_h2_outgas}, high and low outgassing rates can be differentiated on rocky WD planets with H$_2$ atmospheres, although inferring tectonic activities of rocky exoplanets from outgassing rates is limited by many uncertainties. Here, we focus on weakly and highly oxidizing atmospheres dominated by N$_2$ or CO$_2$ and quantitatively discuss the ability of JWST to distinguish between high and low outgassing scenarios for the N$_2$- and CO$_2$-dominated models.

Transmission spectra can differentiate between our high and low outgassing scenarios assuming N$_2$ or CO$_2$ atmospheres, albeit more telescope time is required due to the smaller extent of high MMW atmospheres. We compare simulated transmission spectra for our high and low outgassing scenarios in Figure \ref{fig:spectra_pandexo_n2} and shows molecular contributions to the overall spectra in Figure \ref{fig:individual_species_n2} for N$_2$-dominated atmospheres. We show the same information for CO$_2$-dominated atmospheres in Figure \ref{fig:spectra_pandexo_co2} and Figure \ref{fig:individual_species_co2}. We model an array of campaign sizes with NIRSpec Prism, including 1, 5, 10, and 25 transits, and choose to show the 25-transit program. This is because programs with 10 or less transits have low SNR due to small scale height of N$_2$ and CO$_2$ atmospheres.

For the N$_2$-dominated models, CH$_4$ features at 1.7, 2.3, and 3.4 $\micron$ are the keys to distinguish between the high and low outgassing scenarios (Figure \ref{fig:spectra_pandexo_n2}). CO$_2$ feature at 4.2 $\micron$ also differs significantly between the two scenarios but is a less effective indicator due to lower SNR. For the 6000 K WD models, CH$_4$ features differ by $\sim 2.5\,\sigma$ between high and low outgassing scenarios. CH$_4$ mixing ratio rapidly increases when outgassing rates are high as effective temperature of the host WD decreases, due to the photochemical runaway mechanism discussed in Section \ref{sec:34_h2_test_runaway}. Therefore, high and low outgassing scenarios can be differentiated with larger significance for the 5000 K and 4000 K WD models.

For the CO$_2$-dominated models, high and low outgassing scenarios are not distinguishable with 25 NIRSpec Prism transit for the 6000 K and 5000 K WD models (Figure \ref{fig:spectra_pandexo_co2}). In an oxidizing environment dominated by CO$_2$, CH$_4$ cannot efficiently accumulate even if the emission flux increases by a factor of 1000, explaining the lack of difference between the two scenarios. For the 4000 K WD models, however, CH$_4$ provides opportunity for distinguishing between high and low outgassing rates, likely because emission of CH$_4$ passes the low runaway threshold on a planet with extremely low UV radiation. Using the CH$_4$ features between 0.9 and 2.5 $\micron$ as indicators, we may differentiate between the two outgassing scenarios at a $\sim 10\,\sigma$ significance with 25 transits, and at $\sim 5\,\sigma$ with 10 transits.

\subsection{Testing Photochemical Runaway in Oxidizing Atmospheres} \label{sec:43_test_runaway}
Redox state impacts photochemical runaway \citep[e.g.,][]{ranjan_photochemical_2022}. The main sink of CH$_4$ is hydroxyl (OH) radicals, which is typically produced by O($^1$D) reacting with water. More oxidizing environments naturally lead to higher production of OH radicals, making the photochemical sink of CH$_4$ harder to saturate. Indeed, the mixing ratio of CH$_4$ is generally much lower in N$_2$- and CO$_2$-dominated models, compared to H$_2$-dominated models with the same outgassing rate and WD host (Table \ref{table:ch4_buildup}).

Nevertheless, CH$_4$ runaway can occur in weakly oxidizing and oxidizing atmospheres, as evidenced by strong CH$_4$ features in the 4000 K WD models for both atmospheric redox states (Figure \ref{fig:spectra_pandexo_n2} and \ref{fig:spectra_pandexo_co2}). As a result of photochemical runaway, CH$_4$ is easily detectable by JWST on rocky exoplanets transiting cool WDs. Here we quantitatively discuss CH$_4$ detectability by NIRSpec Prism in our various models.

Detecting CH$_4$ is more challenging in N$_2$-dominated atmospheres compared to H$_2$-dominated atmospheres but is still achievable within a small JWST program for the high outgassing scenarios. For the 4000 K WD model, which has the highest CH$_4$ mixing ratios due to low UV radiation, 5 NIRSpec Prism transits can detect CH$_4$ features at $\sim3$--$4\,\sigma$. For the N$_2$ 5000 K WD and N$_2$ 6000 K WD high outgassing models, detecting CH$_4$ is more complicated -- CH$_4$ features at $< 2\,\micron$ partially or totally overlap with H$_2$O features with similar strength, while CH$_4$ features at $> 2\,\micron$ suffer from higher noise levels. Therefore, confidently detecting CH$_4$ in N$_2$-dominated atmospheres with high outgassing rates based on the 1.7 and 2.3 $\micron$ features requires 25 transits. For all the low outgassing scenarios, CH$_4$ is not detectable.

For the CO$_2$-dominated models, CH$_4$ only reach detectable levels via photochemical runaway in the 4000 K high outgassing scenario. In this case, 5 NIRSpec Prism transits can detect CH$_4$ at $\sim3$--$4\,\sigma$ and 10 transits can achieve a conclusive $\sim 5\,\sigma$ detection.

\subsection{Detectability of Spectral Features in N$_2$- and CO$_2$-dominated Atmospheres} \label{sec:44_detectability}

Here we summarize the locations and detectability of key spectral features in N$_2$- and CO$_2$-dominated atmospheres based on simulated transmission spectra and JWST observations (Figure \ref{fig:spectra_pandexo_n2} for N$_2$ atmospheres and Figure \ref{fig:spectra_pandexo_co2} for CO$_2$ atmospheres). We also show the molecular contributions to the overall spectra (Figure \ref{fig:individual_species_n2} for N$_2$ atmospheres and Figure \ref{fig:individual_species_co2} for CO$_2$ atmospheres). Note that in Figure \ref{fig:spectra_pandexo_h2} we show JWST observations with 1 NIRSpec Prism transit for the H$_2$ models, but for N$_2$- and CO$_2$-dominated models we show 25 transits, because high MMW N$_2$ and CO$_2$ atmospheres have lower atmospheric scale heights. Also note that we assume the stellar parameters of WD 1856+534 \citep{vanderburg_giant_2020}. A brighter or more close-by host can increase SNR and hence detectability of spectral features.

In N$_2$-dominated atmosphere models, H$_2$O have strong features in all scenarios. The strongest water features are located at 2.6 and 6.4 $\micron$, with a few weaker features located at approximately 1.0, 1.2, 1.3, and 1.9 $\micron$, as well as the broad continuum from about 10 to 20 $\micron$. Water features at 1.0-2.6 $\micron$ coincide with a bandpass in which noise levels are low and are likely detectable by NIRSpec Prism. CO$_2$ features at 4.2 and 15 $\micron$ are present but weaker than water features in the reduced outgassing models, but are predominant in the high outgassing models, where CO$_2$ outgassing rates are 1000 times higher than in the reduced outgassing models. In the high outgassing models, CH$_4$ also shows strong features at 3.4 and 7.5 $\micron$, as well as several weaker features in the visible wavelength range. As discussed in Section \ref{sec:43_test_runaway}, these CH$_4$ features are detectable by NIRSpec Prism at high significance and can potentially differentiate between the high and low outgassing scenarios. In addition, N$_2$-N$_2$ has a CIA feature at 4.3 $\micron$, which overlaps with the 4.2 $\micron$ CO$_2$ feature. In the reduced outgassing models, the N$_2$-N$_2$ CIA and the CO$_2$ feature have similar strength and may lead to detection degeneracy.

In a CO$_2$-dominated atmosphere (90\% CO$_2$), CO$_2$ features unsurprisingly dominate the transmission spectra. The strongest CO$_2$ feature is at 15 $\micron$, with broad wings extending to about 11 and 20 $\micron$ on each side. The second strongest CO$_2$ feature is located at 4.2 $\micron$ and is likely the most easily detectable feature for NIRSpec Prism. In addition, CO$_2$ has a few weaker features at 1.4, 1.6, 1.9, 2.1, 2.8, 3.0, 4.8, 5.2, and 10.3 $\micron$. Other than single-molecule absorption features, a CO$_2$-CO$_2$ CIA feature is also present at about 7.5 $\micron$. CO has a feature at about 4.7 $\micron$ but is obscured by the stronger CO$_2$ feature at 4.8 $\micron$. H$_2$O has a broad feature at 6.4 $\micron$, which is the strongest feature of water and does not overlap with any other features. H$_2$O also has a continuum from about 10 to 20 $\micron$, several weak features in visible wavelengths and a strong feature at 2.6 $\micron$, but all these features are obscured by nearby CO$_2$ features. CH$_4$ has two broad features at 3.4 and 7.5 $\micron$, as well as several features in the visible and near-IR wavelengths spanning from 0.9 to 2.5 $\micron$. CH$_4$ features are only strong enough to be detectable in the 5000 K and 4000 K high outgassing cases. Especially in the 4000 K high outgassing scenario, CH$_4$ features have comparable strength as the strongest CO$_2$ features and dominate the visible to near-IR wavelength range probed by NIRSpec Prism.

% \section{Using Chinese, Japanese, and Korean characters}

% Authors have the option to include names in Chinese, Japanese, or Korean (CJK) 
% characters in addition to the English name. The names will be displayed 
% in parentheses after the English name. The way to do this in AASTeX is to 
% use the CJK package available at \url{https://ctan.org/pkg/cjk?lang=en}.
% Further details on how to implement this and solutions for common problems,
% please go to \url{https://journals.aas.org/nonroman/}.

% pdflatex main.tex
% bibtext main
% pdflatex main.tex
% pdflatex main.tex

\bibliography{main}{}

\begin{thebibliography}{}
\expandafter\ifx\csname natexlab\endcsname\relax\def\natexlab#1{#1}\fi
\providecommand{\url}[1]{\href{#1}{#1}}
\providecommand{\dodoi}[1]{doi:~\href{http://doi.org/#1}{\nolinkurl{#1}}}
\providecommand{\doeprint}[1]{\href{http://ascl.net/#1}{\nolinkurl{http://ascl.net/#1}}}
\providecommand{\doarXiv}[1]{\href{https://arxiv.org/abs/#1}{\nolinkurl{https://arxiv.org/abs/#1}}}

\bibitem[{Agol(2011)}]{agol_transit_2011}
Agol, E. 2011, The Astrophysical Journal, 731, L31,
  \dodoi{10.1088/2041-8205/731/2/L31}

\bibitem[{{Airapetian} {et~al.}(2017){Airapetian}, {Glocer}, {Khazanov},
  {Loyd}, {France}, {Sojka}, {Danchi}, \&
  {Liemohn}}]{Airapetian_2017ApJ...836L...3A}
{Airapetian}, V.~S., {Glocer}, A., {Khazanov}, G.~V., {et~al.} 2017, \apjl,
  836, L3, \dodoi{10.3847/2041-8213/836/1/L3}

\bibitem[{Arney {et~al.}(2018)Arney, Domagal-Goldman, \&
  Meadows}]{arney_organic_2018}
Arney, G., Domagal-Goldman, S.~D., \& Meadows, V.~S. 2018, Astrobiology, 18,
  311, \dodoi{10.1089/ast.2017.1666}

\bibitem[{Arney {et~al.}(2016)Arney, Domagal-Goldman, Meadows, Wolf,
  Schwieterman, Charnay, Claire, Hébrard, \& Trainer}]{arney_pale_2016}
Arney, G., Domagal-Goldman, S.~D., Meadows, V.~S., {et~al.} 2016, Astrobiology,
  16, 873, \dodoi{10.1089/ast.2015.1422}

\bibitem[{Arney {et~al.}(2017)Arney, Meadows, Domagal-Goldman, Deming,
  Robinson, Tovar, Wolf, \& Schwieterman}]{arney_pale_2017}
Arney, G.~N., Meadows, V.~S., Domagal-Goldman, S.~D., {et~al.} 2017, The
  Astrophysical Journal, 836, 49, \dodoi{10.3847/1538-4357/836/1/49}

\bibitem[{Batalha {et~al.}(2017)Batalha, Mandell, Pontoppidan, Stevenson,
  Lewis, Kalirai, Earl, Greene, Albert, \& Nielsen}]{batalha_pandexo_2017}
Batalha, N.~E., Mandell, A., Pontoppidan, K., {et~al.} 2017, Publications of
  the Astronomical Society of the Pacific, 129, 064501,
  \dodoi{10.1088/1538-3873/aa65b0}

\bibitem[{Bear \& Soker(2015)}]{bear_planetary_2015}
Bear, E., \& Soker, N. 2015, Monthly Notices of the Royal Astronomical Society,
  450, 4233, \dodoi{10.1093/mnras/stv921}

\bibitem[{{Bergeron} {et~al.}(2001){Bergeron}, {Leggett}, \&
  {Ruiz}}]{Bergeron_2001ApJS..133..413B}
{Bergeron}, P., {Leggett}, S.~K., \& {Ruiz}, M.~T. 2001, \apjs, 133, 413,
  \dodoi{10.1086/320356}

\bibitem[{{B{\'e}tr{\'e}mieux} \&
  {Kaltenegger}(2014)}]{betremieux_2014ApJ...791....7B}
{B{\'e}tr{\'e}mieux}, Y., \& {Kaltenegger}, L. 2014, \apj, 791, 7,
  \dodoi{10.1088/0004-637X/791/1/7}

\bibitem[{{Blackman} {et~al.}(2021){Blackman}, {Beaulieu}, {Bennett},
  {Danielski}, {Alard}, {Cole}, {Vandorou}, {Ranc}, {Terry}, {Bhattacharya},
  {Bond}, {Bachelet}, {Veras}, {Koshimoto}, {Batista}, \&
  {Marquette}}]{Blackman_2021Natur.598..272B}
{Blackman}, J.~W., {Beaulieu}, J.~P., {Bennett}, D.~P., {et~al.} 2021, \nat,
  598, 272, \dodoi{10.1038/s41586-021-03869-6}

\bibitem[{Catling \& Kasting(2017)}]{catling_kasting_2017}
Catling, D.~C., \& Kasting, J.~F. 2017, Atmospheric Evolution on Inhabited and
  Lifeless Worlds (Cambridge University Press), \dodoi{10.1017/9781139020558}

\bibitem[{Chen {et~al.}(2019)Chen, Wolf, Zhan, \&
  Horton}]{chen_habitability_2019}
Chen, H., Wolf, E.~T., Zhan, Z., \& Horton, D.~E. 2019, The Astrophysical
  Journal, 886, 16, \dodoi{10.3847/1538-4357/ab4f7e}

\bibitem[{Cortés \& Kipping(2019)}]{cortes_detectability_2019}
Cortés, J., \& Kipping, D. 2019, Monthly Notices of the Royal Astronomical
  Society, 488, 1695, \dodoi{10.1093/mnras/stz1300}

\bibitem[{{Coutu} {et~al.}(2019){Coutu}, {Dufour}, {Bergeron}, {Blouin},
  {Loranger}, {Allard}, \& {Dunlap}}]{Coutu_2019ApJ...885...74C}
{Coutu}, S., {Dufour}, P., {Bergeron}, P., {et~al.} 2019, \apj, 885, 74,
  \dodoi{10.3847/1538-4357/ab46b9}

\bibitem[{{Davies} \& {Davies}(2010)}]{Davies_2010SolE....1....5D}
{Davies}, J.~H., \& {Davies}, D.~R. 2010, Solid Earth, 1, 5,
  \dodoi{10.5194/se-1-5-2010}

\bibitem[{{de Pater} \& {Lissauer}(2001)}]{dePater_2001plsc.book.....D}
{de Pater}, I., \& {Lissauer}, J.~J. 2001, {Planetary Sciences}

\bibitem[{{Dong} {et~al.}(2017){Dong}, {Huang}, {Lingam}, {T{\'o}th},
  {Gombosi}, \& {Bhattacharjee}}]{Dong_2017ApJ...847L...4D}
{Dong}, C., {Huang}, Z., {Lingam}, M., {et~al.} 2017, \apjl, 847, L4,
  \dodoi{10.3847/2041-8213/aa8a60}

\bibitem[{Dorn {et~al.}(2018)Dorn, Noack, \& Rozel}]{dorn_outgassing_2018}
Dorn, C., Noack, L., \& Rozel, A.~B. 2018, Astronomy \& Astrophysics, 614, A18,
  \dodoi{10.1051/0004-6361/201731513}

\bibitem[{Doyle {et~al.}(2019)Doyle, Young, Klein, Zuckerman, \&
  Schlichting}]{doyle_oxygen_2019}
Doyle, A.~E., Young, E.~D., Klein, B., Zuckerman, B., \& Schlichting, H.~E.
  2019, Science, 366, 356, \dodoi{10.1126/science.aax3901}

\bibitem[{Elkins‐Tanton \& Seager(2008)}]{elkinstanton_ranges_2008}
Elkins‐Tanton, L.~T., \& Seager, S. 2008, The Astrophysical Journal, 685,
  1237, \dodoi{10.1086/591433}

\bibitem[{{Farihi} {et~al.}(2016){Farihi}, {Koester}, {Zuckerman}, {Vican},
  {G{\"a}nsicke}, {Smith}, {Walth}, \& {Breedt}}]{Farihi_2016MNRAS.463.3186F}
{Farihi}, J., {Koester}, D., {Zuckerman}, B., {et~al.} 2016, \mnras, 463, 3186,
  \dodoi{10.1093/mnras/stw2182}

\bibitem[{Fischer(2008)}]{fischer_fluxes_2008}
Fischer, T.~P. 2008, GEOCHEMICAL JOURNAL, 42, 21,
  \dodoi{10.2343/geochemj.42.21}

\bibitem[{Foley \& Smye(2018)}]{foley_carbon_2018}
Foley, B.~J., \& Smye, A.~J. 2018, Astrobiology, 18, 873,
  \dodoi{10.1089/ast.2017.1695}

\bibitem[{{Fontaine} \& {Brassard}(2008)}]{Fontaine_2008PASP..120.1043F}
{Fontaine}, G., \& {Brassard}, P. 2008, \pasp, 120, 1043,
  \dodoi{10.1086/592788}

\bibitem[{Fontaine {et~al.}(2001)Fontaine, Brassard, \&
  Bergeron}]{fontaine_potential_2001}
Fontaine, G., Brassard, P., \& Bergeron, P. 2001, Publications of the
  Astronomical Society of the Pacific, 113, 409, \dodoi{10.1086/319535}

\bibitem[{Fossati {et~al.}(2012)Fossati, Bagnulo, Haswell, Patel, Busuttil,
  Kowalski, Shulyak, \& Sterzik}]{fossati_habitability_2012}
Fossati, L., Bagnulo, S., Haswell, C.~A., {et~al.} 2012, The Astrophysical
  Journal, 757, L15, \dodoi{10.1088/2041-8205/757/1/L15}

\bibitem[{Gaillard \& Scaillet(2014)}]{gaillard_theoretical_2014}
Gaillard, F., \& Scaillet, B. 2014, Earth and Planetary Science Letters, 403,
  307, \dodoi{10.1016/j.epsl.2014.07.009}

\bibitem[{Gillmann \& Tackley(2014)}]{gillmann_atmospheremantle_2014}
Gillmann, C., \& Tackley, P. 2014, Journal of Geophysical Research: Planets,
  119, 1189, \dodoi{10.1002/2013JE004505}

\bibitem[{Grimm \& Heng(2015)}]{grimm_helios-k_2015}
Grimm, S.~L., \& Heng, K. 2015, The Astrophysical Journal, 808, 182,
  \dodoi{10.1088/0004-637X/808/2/182}

\bibitem[{Guillot(2010)}]{guillot_radiative_2010}
Guillot, T. 2010, Astronomy and Astrophysics, 520, A27,
  \dodoi{10.1051/0004-6361/200913396}

\bibitem[{Harman {et~al.}(2021)Harman, Kopparapu, Stefánsson, Lin, Mahadevan,
  Hedges, \& Batalha}]{harman_snowball_2021}
Harman, C.~E., Kopparapu, R.~K., Stefánsson, G., {et~al.} 2021,
  arXiv:2109.10838 [astro-ph].
\newblock \url{http://arxiv.org/abs/2109.10838}

\bibitem[{{Holland}(1984)}]{Holland_1984ceao.book.....H}
{Holland}, H.~D. 1984, {The chemical evolution of the atmosphere and oceans.}

\bibitem[{Hu {et~al.}(2012)Hu, Seager, \& Bains}]{hu_photochemistry_2012}
Hu, R., Seager, S., \& Bains, W. 2012, The Astrophysical Journal, 761, 166,
  \dodoi{10.1088/0004-637X/761/2/166}

\bibitem[{{Huang} {et~al.}(2021){Huang}, {Seager}, {Petkowski}, {Ranjan}, \&
  {Zhan}}]{Huang_2021arXiv210712424H}
{Huang}, J., {Seager}, S., {Petkowski}, J.~J., {Ranjan}, S., \& {Zhan}, Z.
  2021, arXiv e-prints, arXiv:2107.12424.
\newblock \doarXiv{2107.12424}

\bibitem[{{Hunten}(1973)}]{Hunten_1973JAtS...30.1481H}
{Hunten}, D.~M. 1973, Journal of Atmospheric Sciences, 30, 1481

\bibitem[{James \& Hu(2018)}]{james_photochemical_2018}
James, T., \& Hu, R. 2018, The Astrophysical Journal, 867, 17,
  \dodoi{10.3847/1538-4357/aae2bb}

\bibitem[{Kasting \& Catling(2003)}]{kasting_evolution_2003}
Kasting, J.~F., \& Catling, D. 2003, Annual Review of Astronomy and
  Astrophysics, 41, 429, \dodoi{10.1146/annurev.astro.41.071601.170049}

\bibitem[{{Kasting} {et~al.}(1993){Kasting}, {Whitmire}, \&
  {Reynolds}}]{Kasting_1993Icar..101..108K}
{Kasting}, J.~F., {Whitmire}, D.~P., \& {Reynolds}, R.~T. 1993, \icarus, 101,
  108, \dodoi{10.1006/icar.1993.1010}

\bibitem[{Keller-Rudek {et~al.}(2013)Keller-Rudek, Moortgat, Sander, \&
  Sörensen}]{keller-rudek_mpi-mainz_2013}
Keller-Rudek, H., Moortgat, G.~K., Sander, R., \& Sörensen, R. 2013, Earth
  System Science Data, 9

\bibitem[{Kepler {et~al.}(2016)Kepler, Pelisoli, Koester, Ourique, Romero,
  Reindl, Kleinman, Eisenstein, Valois, \& Amaral}]{kepler_new_2016}
Kepler, S.~O., Pelisoli, I., Koester, D., {et~al.} 2016, Monthly Notices of the
  Royal Astronomical Society, 455, 3413, \dodoi{10.1093/mnras/stv2526}

\bibitem[{Kite {et~al.}(2019)Kite, {Bruce Fegley Jr.}, Schaefer, \&
  Ford}]{kite_superabundance_2019}
Kite, E.~S., {Bruce Fegley Jr.}, Schaefer, L., \& Ford, E.~B. 2019, The
  Astrophysical Journal, 887, L33, \dodoi{10.3847/2041-8213/ab59d9}

\bibitem[{Kite {et~al.}(2020)Kite, Fegley~Jr., Schaefer, \&
  Ford}]{kite_atmosphere_2020}
Kite, E.~S., Fegley~Jr., B., Schaefer, L., \& Ford, E.~B. 2020, The
  Astrophysical Journal, 891, 111, \dodoi{10.3847/1538-4357/ab6ffb}

\bibitem[{{Komacek} {et~al.}(2020){Komacek}, {Fauchez}, {Wolf}, \&
  {Abbot}}]{Komacek_2020ApJ...888L..20K}
{Komacek}, T.~D., {Fauchez}, T.~J., {Wolf}, E.~T., \& {Abbot}, D.~S. 2020,
  \apjl, 888, L20, \dodoi{10.3847/2041-8213/ab6200}

\bibitem[{Kozakis \& Kaltenegger(2019)}]{kozakis_atmospheres_2019}
Kozakis, T., \& Kaltenegger, L. 2019, The Astrophysical Journal, 875, 99,
  \dodoi{10.3847/1538-4357/ab11d3}

\bibitem[{Kozakis {et~al.}(2018)Kozakis, Kaltenegger, \&
  Hoard}]{kozakis_uv_2018}
Kozakis, T., Kaltenegger, L., \& Hoard, D.~W. 2018, The Astrophysical Journal,
  862, 69, \dodoi{10.3847/1538-4357/aacbc7}

\bibitem[{Kozakis {et~al.}(2020)Kozakis, Lin, \&
  Kaltenegger}]{kozakis_high-resolution_2020}
Kozakis, T., Lin, Z., \& Kaltenegger, L. 2020, The Astrophysical Journal, 894,
  L6, \dodoi{10.3847/2041-8213/ab6f6a}

\bibitem[{Kulikov {et~al.}(2007)Kulikov, Lammer, Lichtenegger, Penz, Breuer,
  Spohn, Lundin, \& Biernat}]{kulikov_comparative_2007}
Kulikov, Y.~N., Lammer, H., Lichtenegger, H. I.~M., {et~al.} 2007, Space
  Science Reviews, 129, 207, \dodoi{10.1007/s11214-007-9192-4}

\bibitem[{{Lederberg}(1965)}]{Lederberg_1965Natur.207....9L}
{Lederberg}, J. 1965, \nat, 207, 9, \dodoi{10.1038/207009a0}

\bibitem[{Lenardic {et~al.}(2016)Lenardic, Jellinek, Foley, O'Neill, \&
  Moore}]{lenardic_climate-tectonic_2016}
Lenardic, A., Jellinek, A.~M., Foley, B., O'Neill, C., \& Moore, W.~B. 2016,
  Journal of Geophysical Research: Planets, 121, 1831,
  \dodoi{10.1002/2016JE005089}

\bibitem[{Lin {et~al.}(2021)Lin, MacDonald, Kaltenegger, \&
  Wilson}]{lin_differentiating_2021}
Lin, Z., MacDonald, R.~J., Kaltenegger, L., \& Wilson, D.~J. 2021, Monthly
  Notices of the Royal Astronomical Society, 505, 3562,
  \dodoi{10.1093/mnras/stab1486}

\bibitem[{Loeb \& Maoz(2013)}]{loeb_detecting_2013}
Loeb, A., \& Maoz, D. 2013, Monthly Notices of the Royal Astronomical Society:
  Letters, 432, L11, \dodoi{10.1093/mnrasl/slt026}

\bibitem[{{Lopez} \& {Fortney}(2013)}]{Lopez_2013ApJ...776....2L}
{Lopez}, E.~D., \& {Fortney}, J.~J. 2013, \apj, 776, 2,
  \dodoi{10.1088/0004-637X/776/1/2}

\bibitem[{Luger \& Barnes(2015)}]{luger_extreme_2015}
Luger, R., \& Barnes, R. 2015, Astrobiology, 15, 119,
  \dodoi{10.1089/ast.2014.1231}

\bibitem[{Malamud \& Perets(2017)}]{malamud_post-main-sequence_2017-1}
Malamud, U., \& Perets, H.~B. 2017, The Astrophysical Journal, 849, 8,
  \dodoi{10.3847/1538-4357/aa8df5}

\bibitem[{{Marley} {et~al.}(2013){Marley}, {Ackerman}, {Cuzzi}, \&
  {Kitzmann}}]{Marley_2013cctp.book..367M}
{Marley}, M.~S., {Ackerman}, A.~S., {Cuzzi}, J.~N., \& {Kitzmann}, D. 2013,
  {Clouds and Hazes in Exoplanet Atmospheres}, ed. S.~J. {Mackwell}, A.~A.
  {Simon-Miller}, J.~W. {Harder}, \& M.~A. {Bullock}, 367,
  \dodoi{10.2458/azu\_uapress\_9780816530595-ch15}

\bibitem[{{Molli{\`e}re} {et~al.}(2019){Molli{\`e}re}, {Wardenier}, {van
  Boekel}, {Henning}, {Molaverdikhani}, \&
  {Snellen}}]{Molliere_2019A&A...627A..67M}
{Molli{\`e}re}, P., {Wardenier}, J.~P., {van Boekel}, R., {et~al.} 2019, \aap,
  627, A67, \dodoi{10.1051/0004-6361/201935470}

\bibitem[{{Ortenzi} {et~al.}(2020){Ortenzi}, {Noack}, {Sohl}, {Guimond},
  {Grenfell}, {Dorn}, {Schmidt}, {Vulpius}, {Katyal}, {Kitzmann}, \&
  {Rauer}}]{Ortenzi_2020NatSR..1010907O}
{Ortenzi}, G., {Noack}, L., {Sohl}, F., {et~al.} 2020, Scientific Reports, 10,
  10907, \dodoi{10.1038/s41598-020-67751-7}

\bibitem[{Owen {et~al.}(2020)Owen, Shaikhislamov, Lammer, Fossati, \&
  Khodachenko}]{owen_hydrogen_2020}
Owen, J.~E., Shaikhislamov, I.~F., Lammer, H., Fossati, L., \& Khodachenko,
  M.~L. 2020, Space Science Reviews, 216, 129,
  \dodoi{10.1007/s11214-020-00756-w}

\bibitem[{Owen \& Wu(2013)}]{owen_kepler_2013}
Owen, J.~E., \& Wu, Y. 2013, The Astrophysical Journal, 12

\bibitem[{{Owen} \& {Wu}(2017)}]{Owen_2017ApJ...847...29O}
{Owen}, J.~E., \& {Wu}, Y. 2017, \apj, 847, 29,
  \dodoi{10.3847/1538-4357/aa890a}

\bibitem[{Perryman {et~al.}(2014)Perryman, Hartman, Bakos, \&
  Lindegren}]{perryman_astrometric_2014}
Perryman, M., Hartman, J., Bakos, G.~A., \& Lindegren, L. 2014, The
  Astrophysical Journal, 797, 14, \dodoi{10.1088/0004-637X/797/1/14}

\bibitem[{Ramirez \& Kaltenegger(2016)}]{ramirez_habitable_2016}
Ramirez, R.~M., \& Kaltenegger, L. 2016, The Astrophysical Journal, 823, 6,
  \dodoi{10.3847/0004-637X/823/1/6}

\bibitem[{Ramirez {et~al.}(2014)Ramirez, Kopparapu, Zugger, Robinson, Freedman,
  \& Kasting}]{ramirez_warming_2014}
Ramirez, R.~M., Kopparapu, R., Zugger, M.~E., {et~al.} 2014, Nature Geoscience,
  7, 59, \dodoi{10.1038/ngeo2000}

\bibitem[{{Ranjan} {et~al.}(2020){Ranjan}, {Schwieterman}, {Harman}, {Fateev},
  {Sousa-Silva}, {Seager}, \& {Hu}}]{Ranjan_2020ApJ...896..148R}
{Ranjan}, S., {Schwieterman}, E.~W., {Harman}, C., {et~al.} 2020, \apj, 896,
  148, \dodoi{10.3847/1538-4357/ab9363}

\bibitem[{Ranjan {et~al.}(2022)Ranjan, Seager, Zhan, Koll, Bains, Petkowski,
  Huang, \& Lin}]{ranjan_photochemical_2022}
Ranjan, S., Seager, S., Zhan, Z., {et~al.} 2022, arXiv:2201.08359 [astro-ph].
\newblock \url{http://arxiv.org/abs/2201.08359}

\bibitem[{{Ranjan} {et~al.}(2018){Ranjan}, {Todd}, {Sutherland}, \&
  {Sasselov}}]{Ranjan_2018AsBio..18.1023R}
{Ranjan}, S., {Todd}, Z.~R., {Sutherland}, J.~D., \& {Sasselov}, D.~D. 2018,
  Astrobiology, 18, 1023, \dodoi{10.1089/ast.2017.1770}

\bibitem[{Rey {et~al.}(2014)Rey, Coltice, \& Flament}]{rey_spreading_2014}
Rey, P.~F., Coltice, N., \& Flament, N. 2014, Nature, 513, 405,
  \dodoi{10.1038/nature13728}

\bibitem[{{Robinson} {et~al.}(2017){Robinson}, {Fortney}, \&
  {Hubbard}}]{Robinson_2017ApJ...850..128R}
{Robinson}, T.~D., {Fortney}, J.~J., \& {Hubbard}, W.~B. 2017, \apj, 850, 128,
  \dodoi{10.3847/1538-4357/aa951e}

\bibitem[{Roble {et~al.}(1987)Roble, Ridley, \& Dickinson}]{roble_global_1987}
Roble, R.~G., Ridley, E.~C., \& Dickinson, R.~E. 1987, Journal of Geophysical
  Research, 92, 8745, \dodoi{10.1029/JA092iA08p08745}

\bibitem[{{Saumon} {et~al.}(2014){Saumon}, {Holberg}, \&
  {Kowalski}}]{Saumon_2014ApJ...790...50S}
{Saumon}, D., {Holberg}, J.~B., \& {Kowalski}, P.~M. 2014, \apj, 790, 50,
  \dodoi{10.1088/0004-637X/790/1/50}

\bibitem[{Scaillet \& Gaillard(2011)}]{scaillet_redox_2011}
Scaillet, B., \& Gaillard, F. 2011, Nature, 480, 48, \dodoi{10.1038/480048a}

\bibitem[{Schaefer \& Fegley(2010)}]{schaefer_chemistry_2010}
Schaefer, L., \& Fegley, B. 2010, Icarus, 208, 438,
  \dodoi{10.1016/j.icarus.2010.01.026}

\bibitem[{{Schreiber} {et~al.}(2019){Schreiber}, {G{\"a}nsicke}, {Toloza},
  {Hernandez}, \& {Lagos}}]{Schreiber_2019ApJ...887L...4S}
{Schreiber}, M.~R., {G{\"a}nsicke}, B.~T., {Toloza}, O., {Hernandez}, M.-S., \&
  {Lagos}, F. 2019, \apjl, 887, L4, \dodoi{10.3847/2041-8213/ab42e2}

\bibitem[{{Schr{\"o}der} \& {Smith}(2008)}]{Schroder_2008MNRAS.386..155S}
{Schr{\"o}der}, K.~P., \& {Smith}, R.~C. 2008, \mnras, 386, 155,
  \dodoi{10.1111/j.1365-2966.2008.13022.x}

\bibitem[{Schwieterman {et~al.}(2019)Schwieterman, Reinhard, Olson, Ozaki,
  Harman, Hong, \& Lyons}]{schwieterman_rethinking_2019}
Schwieterman, E.~W., Reinhard, C.~T., Olson, S.~L., {et~al.} 2019, The
  Astrophysical Journal, 874, 9, \dodoi{10.3847/1538-4357/ab05e1}

\bibitem[{Seager {et~al.}(2013)Seager, Bains, \& Hu}]{seager_biosignature_2013}
Seager, S., Bains, W., \& Hu, R. 2013, The Astrophysical Journal, 777, 95,
  \dodoi{10.1088/0004-637X/777/2/95}

\bibitem[{{Segura} {et~al.}(2005){Segura}, {Kasting}, {Meadows}, {Cohen},
  {Scalo}, {Crisp}, {Butler}, \& {Tinetti}}]{Segura_2005AsBio...5..706S}
{Segura}, A., {Kasting}, J.~F., {Meadows}, V., {et~al.} 2005, Astrobiology, 5,
  706, \dodoi{10.1089/ast.2005.5.706}

\bibitem[{Sengupta(2016)}]{sengupta_upper_2016}
Sengupta, S. 2016, Journal of Astrophysics and Astronomy, 37, 11,
  \dodoi{10.1007/s12036-016-9390-0}

\bibitem[{Sousa-Silva {et~al.}(2020)Sousa-Silva, Seager, Ranjan, Petkowski,
  Zhan, Hu, \& Bains}]{sousa-silva_phosphine_2020}
Sousa-Silva, C., Seager, S., Ranjan, S., {et~al.} 2020, Astrobiology, 20, 235,
  \dodoi{10.1089/ast.2018.1954}

\bibitem[{{Stephan} {et~al.}(2017){Stephan}, {Naoz}, \&
  {Zuckerman}}]{Stephan_2017ApJ...844L..16S}
{Stephan}, A.~P., {Naoz}, S., \& {Zuckerman}, B. 2017, \apjl, 844, L16,
  \dodoi{10.3847/2041-8213/aa7cf3}

\bibitem[{{Suissa} {et~al.}(2020){Suissa}, {Mandell}, {Wolf}, {Villanueva},
  {Fauchez}, \& {Kopparapu}}]{Suissa_2020ApJ...891...58S}
{Suissa}, G., {Mandell}, A.~M., {Wolf}, E.~T., {et~al.} 2020, \apj, 891, 58,
  \dodoi{10.3847/1538-4357/ab72f9}

\bibitem[{Tian {et~al.}(2008)Tian, Kasting, Liu, \&
  Roble}]{tian_hydrodynamic_2008}
Tian, F., Kasting, J.~F., Liu, H.-L., \& Roble, R.~G. 2008, Journal of
  Geophysical Research, 113, E05008, \dodoi{10.1029/2007JE002946}

\bibitem[{Trail {et~al.}(2011)Trail, Watson, \& Tailby}]{trail_oxidation_2011}
Trail, D., Watson, E.~B., \& Tailby, N.~D. 2011, Nature, 480, 79,
  \dodoi{10.1038/nature10655}

\bibitem[{van Sluijs \& Van~Eylen(2018)}]{van_sluijs_occurrence_2018}
van Sluijs, L., \& Van~Eylen, V. 2018, Monthly Notices of the Royal
  Astronomical Society, 474, 4603, \dodoi{10.1093/mnras/stx3068}

\bibitem[{Vanderburg {et~al.}(2020)Vanderburg, Rappaport, Xu, Crossfield,
  Becker, Gary, Murgas, Blouin, Kaye, Palle, Melis, Morris, Kreidberg, Gorjian,
  Morley, Mann, Parviainen, Pearce, Newton, Carrillo, Zuckerman, Nelson,
  Zeimann, Brown, Tronsgaard, Klein, Ricker, Vanderspek, Latham, Seager, Winn,
  Jenkins, Adams, Benneke, Berardo, Buchhave, Caldwell, Christiansen, Collins,
  Colón, Daylan, Doty, Doyle, Dragomir, Dressing, Dufour, Fukui, Glidden,
  Guerrero, Guo, Heng, Henriksen, Huang, Kaltenegger, Kane, Lewis, Lissauer,
  Morales, Narita, Pepper, Rose, Smith, Stassun, \& Yu}]{vanderburg_giant_2020}
Vanderburg, A., Rappaport, S.~A., Xu, S., {et~al.} 2020, Nature, 585, 363,
  \dodoi{10.1038/s41586-020-2713-y}

\bibitem[{van Lieshout {et~al.}(2018)van Lieshout, Kral, Charnoz, Wyatt, \&
  Shannon}]{vanlieshout_exoplanet_2018}
van Lieshout, R., Kral, Q., Charnoz, S., Wyatt, M.~C., \& Shannon, A. 2018,
  Monthly Notices of the Royal Astronomical Society, 480, 2784,
  \dodoi{10.1093/mnras/sty1271}

\bibitem[{Veras \& Gänsicke(2015)}]{veras_detectable_2015}
Veras, D., \& Gänsicke, B.~T. 2015, Monthly Notices of the Royal Astronomical
  Society, 447, 1049, \dodoi{10.1093/mnras/stu2475}

\bibitem[{{Veras} {et~al.}(2014){Veras}, {Shannon}, \&
  {G{\"a}nsicke}}]{Veras_2014MNRAS.445.4175V}
{Veras}, D., {Shannon}, A., \& {G{\"a}nsicke}, B.~T. 2014, \mnras, 445, 4175,
  \dodoi{10.1093/mnras/stu2026}

\bibitem[{Watson {et~al.}(1981)Watson, Donahue, \&
  Walker}]{watson_dynamics_1981}
Watson, A.~J., Donahue, T.~M., \& Walker, J.~C. 1981, Icarus, 48, 150,
  \dodoi{10.1016/0019-1035(81)90101-9}

\bibitem[{Weller {et~al.}(2015)Weller, Lenardic, \&
  O'Neill}]{weller_effects_2015}
Weller, M., Lenardic, A., \& O'Neill, C. 2015, Earth and Planetary Science
  Letters, 420, 85, \dodoi{10.1016/j.epsl.2015.03.021}

\bibitem[{Wood {et~al.}(2006)Wood, Walter, \& Wade}]{wood_accretion_2006}
Wood, B.~J., Walter, M.~J., \& Wade, J. 2006, Nature, 441, 825,
  \dodoi{10.1038/nature04763}

\bibitem[{{Yang} {et~al.}(2014){Yang}, {Bou{\'e}}, {Fabrycky}, \&
  {Abbot}}]{Yang_2014ApJ...787L...2Y}
{Yang}, J., {Bou{\'e}}, G., {Fabrycky}, D.~C., \& {Abbot}, D.~S. 2014, \apjl,
  787, L2, \dodoi{10.1088/2041-8205/787/1/L2}

\bibitem[{{Zhan} {et~al.}(2021){Zhan}, {Seager}, {Petkowski}, {Sousa-Silva},
  {Ranjan}, {Huang}, \& {Bains}}]{Zhan_2021AsBio..21..765Z}
{Zhan}, Z., {Seager}, S., {Petkowski}, J.~J., {et~al.} 2021, Astrobiology, 21,
  765, \dodoi{10.1089/ast.2019.2146}

\end{thebibliography}
\bibliographystyle{aasjournal}



\end{document}